\documentclass[reprint,amsmath,amssymb,aps,prd]{revtex4-2}

\usepackage[colorlinks,linkcolor=red,urlcolor=blue,citecolor=blue]{hyperref}
\usepackage{amssymb, amsmath, bm, dcolumn, epsf, graphicx, latexsym, mathbbol, slashed}
\usepackage{mathrsfs}
\usepackage{comment}
\usepackage{color}
\usepackage{multirow}
\usepackage{array}
\usepackage{subcaption}
\usepackage{caption}
\usepackage[normalem]{ulem}
\usepackage{color}
\newcommand{\dd}[0]{\mathrm{d}}
\allowdisplaybreaks

\begin{document}

\title{Gravitational Lensing of Gravitational Waves: \\Spin-wave Optics through Black Hole Scattering}

\author{Zhao Li}
\email{lz111301@pku.edu.cn}
\affiliation{Department of Astronomy, Peking University, Beijing, 100871, China}
\affiliation{Department of Astronomy, University of Science and Technology of China, Hefei, Anhui 230026, China, and\\ School of Astronomy and Space Science, University of Science and Technology of China, Hefei, Anhui 230026, China}

\author{Shaoqi Hou}
\affiliation{School of Physics and Technology, Wuhan University, Wuhan, Hubei 430072, China}

\author{Wen Zhao}
\email{wzhao7@ustc.edu.cn}
\affiliation{Department of Astronomy, University of Science and Technology of China, Hefei, Anhui 230026, China, and\\ School of Astronomy and Space Science, University of Science and Technology of China, Hefei, Anhui 230026, China}
\affiliation{College of Physics, Guizhou University, Guiyang, Guizhou 550025, China}

\begin{abstract}
Gravitational-wave (GW) scattering in strong gravitational fields is a central problem in GW lensing. Yet, conventional treatments based on asymptotic expansions suffer from divergences and become unreliable near the optical axis. In this work, we present a rigorous calculation of GW scattering by a Schwarzschild black hole (BH) within the BH perturbation theory. By placing the observer at a finite distance and abandoning the asymptotic expansion of radial wave functions, we obtain a well-convergent partial-wave description without invoking any regularization scheme, thereby naturally resolving the divergences of the partial-wave series and the Poisson spot. We numerically computed the scattered GW waveforms by reconstructing the physical $+$ and $\times$ polarizations from the master variables, revealing the formation of the Poisson spot and pronounced wavefront distortions. 
A systematic comparison with conventional asymptotic approaches shows that they reproduce only qualitative features at large scattering angles and fail in the forward-scattering region. We further compare the frequency-domain transmission factors derived from the scattering formalism with those obtained from the Kirchhoff diffraction integral, finding significant discrepancies at high frequencies due to the latter's neglect of long-range gravitational effects and polarization evolution. Our results establish a stable and physically transparent framework for GW scattering in strong-field regimes and provide a solid foundation for accurate modeling of GW lensing beyond standard approximations.
\end{abstract}

\maketitle

\section{\label{introduction}Introduction}

Gravitational lensing is one of the central predictions of general relativity and provides powerful tools for cosmology, including probing high-redshift objects \cite{Dan_2012,Vieira_2013,Yang_2022}, exploring large-scale cosmic structures \cite{Refregier_2003,Cao_2012}, measuring the masses of lensing objects \cite{Bolton_2008,Hou_2019_b}, and testing theories of gravity \cite{Wang_2024,Liu_2022}. When the gravitational wave (GW) passes by a massive object, it is deflected and can form multiple images, in close analogy with electromagnetic lensing \cite{Piorkowska_2013,Lin_2023,Haris_2018,Hannuksela_2019,McIsaac_2020,XiaoshuLiu_2021,LIGO_2021,Lo_2023,Abbott_2024,Lo_2025}, thereby encoding rich astrophysical and theoretical information. Accurate searches for lensed GW signals and reliable parameter estimation critically rely on precise waveform templates, which motivates further theoretical investigations of GW lensing.

The lowest-order description of GW lensing is provided by geometric optics \cite{Isaacson_1968,Keeton_2005,Sereno_2006,Bozza_2010,Hou_2019,Li_2025}. This framework successfully accounts for path deflection, magnification, and the formation of multiple images. However, it fails to capture wave and polarization effects, leading to degeneracies between source and lens parameters \cite{Ezquiaga_2021}. Wave effects become important when the GW wavelength is comparable to the characteristic scale of the lens \cite{Meena_2019,Shan_2025,Yuan_2025}. A particularly relevant class of events for space-based detectors involves GWs emitted by supermassive black hole (BH) coalescences and lensed by foreground supermassive BHs. Recent analyses suggest that, in a wave-optics search, the event GW231123 is more likely to be lensed by an isolated stellar-mass BH, or stellar-mass black holes embedded in a macroscopic gravitational field \cite{LVK_lensing,Chan_2025_b,Goyal_2025,Chakraborty_2025,Shan_2025_b}.
To incorporate wave-optic effects, the Kirchhoff integral approach has been developed, although it still relies on geometric-optic assumptions and neglects the polarization evolution of GW \cite{Nakamura_1998,Baraldo_1999,Takahashi_2003,Guo_2020}.

From the first principle, BH perturbation theory \cite{Regge_Wheeler_1957,Zerilli_1970a,Zerilli_1970b,Teukolsky_1973,Press_1973,Bardeen_1973,Teukolsky_1974,Chandrasekhar_1978a,Chandrasekhar_1978b,Sasaki_Nakamura_1982}, combined with scattering theory \cite{Matzner_1968,Fabbri_1975,Chrzanowski_1976,Sanchez_1978a,Sanchez_1978b,Sanchez_1976,Handler_Matzner_1980,Andersson_1995,Glampedakis_2001,Dolan_2008a,Dolan_2008b,Bao_2022}, provides a more complete framework that captures both wave and polarization effects. Previous studies of BH scattering, such as Refs.\,\cite{Dolan_2008a,Dolan_2008b}, primarily focused on the asymptotic behavior of the scattered waves. As a result, the divergences of the partial-wave series and of the Poisson spot are not properly addressed. This limitation necessitates the introduction of regularization schemes \cite{Yennie_1954,Stratton_2020,Folacci_2019a,Folacci_2019b,Pijnenburg_2024a} to control these divergences at large scattering angles, while still preventing a reliable analysis of the scattered wave in the vicinity of the optical axis \cite{Pijnenburg_2024b,Chan_2025,Saketh_2025,Zhang_Fan_2021,Li_2025_b}.

In this work, we aim to fill this gap by presenting a rigorous calculation of GW scattering in a Schwarzschild spacetime. In contrast to conventional treatments, we discard the asymptotic expansion of the radial wave functions and place the observer at a finite distance, thereby avoiding the divergences of the partial-wave series and the divergence associated with the Poisson spot. Using the numerical method, we present the wave field of scattered GWs and perform two main comparisons between our results and those of the previous approaches. 

First, we compare the diffraction patterns, namely the angular distributions of the scattered GW amplitudes, obtained from our calculation and from conventional treatments using asymptotic expansions, 
The latter reproduces only the qualitative feature of scattered waves in the far-axis region at large scattering angles and leads to an unphysical divergence in the forward-scattering region. Second, we compare the frequency-domain transmission factors, defined as the ratio of lensed to unlensed waveforms, obtained from the scattering formalism and from the Kirchhoff integral, finding significant discrepancies in the high-frequency regime. This results from the fact that the latter treats GWs as massless scalar fields and neglects strong-field effects in BH scattering.

This paper is organized as follows. Section\,\ref{sec:black_hole_perturbation_theory} briefly reviews the essentials of BH perturbation theory. The incident plane GW and the boundary conditions for the scattering problem are discussed in Sec.\,\ref{sec:planar_waves}. The computational procedure for the scattered GWs and their polarizations is presented in Sec.\,\ref{sec:scattered_GW}. The main numerical results and comparisons are given in Sec.\,\ref{sec:numerical_results}. Throughout this work, we use geometric units with $c=G=1$, where $c$ is the speed of light and $G$ is the gravitational constant. The Schwarzschild coordinates are denoted by $x^{\mu}\equiv\{t,r,\theta,\varphi\}$, while the Cartesian coordinates are denoted by $\hat{x}^{\mu}\equiv\{t,x,y,z\}$.

\section{\label{sec:black_hole_perturbation_theory}Black Hole Perturbation theory}

In the linear perturbation theory, the full metric $g_{\mu\nu}$ is decomposed into a background part and a perturbation, $g_{\mu\nu}=\bar{g}_{\mu\nu}+h_{\mu\nu}$.
In this work, the background is taken to be a Schwarzschild BH, whose line element $\dd{s}^2=\bar{g}_{\mu\nu}\dd{x}^{\mu}\dd{x}^{\nu}$ is given by \cite{Weinberg}
\begin{equation}
\dd{s}^2=-f(r)\dd{t}^2+f^{-1}(r)\dd{r}^2+r^2(\dd{\theta}^2+\sin^2\theta\dd{\varphi}^2),
\end{equation}
where $f(r)=1-2M/r$, and $M$ denotes the BH mass.
The gravitational perturbation is governed by the linearized Einstein field equation, $R^{\rm(GW)}_{\mu\nu}=0$, where $R^{\rm(GW)}_{\mu\nu}$ is the Ricci tensor constructed from $h_{\mu\nu}$ \cite{MTW}. 
In a spherically symmetric spacetime, it is convenient to decompose the gravitational perturbation into frequency and angular modes \cite{Maggiore_2018_b},
\begin{equation}
\label{eq_2:Fourier}
h_{\mu\nu}(t,\bm{r})
=\frac{1}{2\pi}\int_{-\infty}^{\infty}\tilde{h}_{\mu\nu}(k,\bm{r})e^{-ikt}\dd k,
\end{equation}
\begin{equation}
\label{eq_2:partial_wave_series}
\tilde{h}_{\mu\nu}(k,\bm{r})
=\sum_{a\ell m}
\tilde{h}^{(a)}_{\ell m}(k,r)
\left[\mathbf{T}_{\ell m}^{(a)}(\theta,\varphi)\right]_{\mu\nu},
\end{equation}
where $k$ is the wave number, $\bm{r}$ denotes the spatial coordinates, and $\ell$ and $m$ are the angular and magnetic quantum numbers.
The tensors $\mathbf{T}_{\ell m}^{(a)}$ form a complete set of tensor harmonics and are eigenstates of angular momentum \cite{Maggiore_2018_a}. Their explicit expressions are listed in Appendix \ref{app:spherical_harmonics}. The superscript $(a)$ labels the ten independent basis tensors,
\begin{equation}
a\in\{tt,Rt,L0,T0,Et,E1,Bt,B1,E2,B2\},
\end{equation}
among which $\{tt, Rt, L0, T0, Et, E1, E2\}$ correspond to parity-even modes, while $\{Bt, B1, B2\}$ correspond to parity-odd modes.
Equation (\ref{eq_2:partial_wave_series}) is referred to as the partial-wave series, with summation ranges $\ell\geqslant0$ for $\{tt, Rt, L0, T0\}$ modes, $\ell\geqslant1$ for $\{Et, E1, Bt, B1\}$ modes, $\ell\geqslant2$ for $\{E2, B2\}$ modes, and $-\ell\leqslant m\leqslant\ell$ for all $\ell$.

In general relativity, the linearized Einstein equation is invariant under gauge transformations,
\begin{equation}
\label{eq_2:gauge_transformation}
h_{\mu\nu}\rightarrow h_{\mu\nu}-(\nabla_{\mu}\xi_{\nu}+\nabla_{\nu}\xi_{\mu}),
\end{equation}
where $\nabla_{\mu}$ is the covariant derivative, compatible with the background spacetime, and can be simplified through an appropriate choice of the gauge vector $\xi_\mu$. 
A commonly adopted choice is the Regge–Wheeler (RW) gauge \cite{Regge_Wheeler_1957}, which preserves the following modes: $Bt$ with $\ell\geqslant 2$, $B1$ with $\ell\geqslant 1$, $tt$ with $\ell\geqslant 0$, $L0$ with $\ell\geqslant 0$, $Rt$ with $\ell\geqslant 1$, and $T0$ with $\ell\geqslant 2$.
Under the RW gauge, the ten metric components in Eq.\,(\ref{eq_2:partial_wave_series}) are reduced to six. In the following, we focus on modes with $\ell\geqslant2$, since lower-$\ell$ modes only enter the constraint equations, which are automatically satisfied. The spherical-harmonic decomposition of the gauge transformation (\ref{eq_2:gauge_transformation}) and the RW gauge conditions are reviewed in Appendix \ref{app:gauge_transformation}.

In general relativity, linear perturbations possess only two independent physical degrees of freedom. Accordingly, two master variables, corresponding to the parity-odd and parity-even sectors, are sufficient to describe the perturbation completely. Following the standard conventions \cite{Regge_Wheeler_1957,Zerilli_1970a,Maggiore_2018_b}, we define the master functions as
\begin{equation}
\label{eq:RW_function}
\tilde{\psi}^{(-)}_{\ell m}=-\frac{f(r)}{r}\tilde{h}^{(B1)}_{\ell m}(r),
\end{equation}
and
\begin{equation}
\label{eq:Zerilli_function}
\tilde{\psi}_{\ell m}^{(+)}\equiv\frac{1}{\Lambda}
\left[\frac{1}{r}\tilde{h}_{\ell m}^{(T0)}
+\frac{f(r)}{ik}
\tilde{h}_{\ell m}^{(Rt)}\right],
\end{equation}
for odd- and even-parity perturbations, respectively. Here, $\Lambda\equiv\lambda+3M/r$, with $\lambda=(\ell-1)(\ell+2)/2$.

The radial functions $\tilde{\psi}_{\ell m}^{(\pm)}$ satisfy two single-variable ordinary differential equations, known as the RW and Zerilli equations,
\begin{equation}
\label{eq_2:radial_equations}
\left[\frac{\dd^2}{\dd r_*^2}+k^2-V^{(\pm)}_{\ell}(r)\right]\tilde{\psi}^{(\pm)}_{\ell m}(k,r)=0,
\end{equation}
with the effective potentials
\begin{equation}
V^{(-)}_{\ell}(r)=\frac{f(r)}{r^2}\left[\ell(\ell+1)-\frac{6M}{r}\right],
\end{equation}
and
\begin{equation}
\begin{aligned}
V^{(+)}_{\ell}(r)&=\frac{f(r)}{r^2}\frac{1}{\Lambda^2}\left[2\lambda^2(\lambda+1)+6\lambda^2\left(\frac{M}{r}\right)\right.\\
&\qquad\left.+18\lambda\left(\frac{M}{r}\right)^2+18\left(\frac{M}{r}\right)^3\right].
\end{aligned}
\end{equation}

For the scattering problem, the radial functions satisfy the asymptotic boundary conditions
\begin{equation}
\label{eq_2:boundary_conditions_inf}
\tilde{\psi}_{\ell m}^{(\pm)}(k,r\rightarrow\infty)\rightarrow
c_{\ell m}^{(\pm)}(k)\left[e^{-ikr_*}-(-1)^{\ell}e^{2i\delta^{(\pm)}_{\ell}}e^{ikr_*}\right].
\end{equation}
at spatial infinity, and
\begin{equation}
\label{eq_2:boundary_conditions_horizon}
\tilde{\psi}_{\ell m}^{(\pm)}(k,r\rightarrow2M)\rightarrow
a_{\ell m}^{(\pm)}(k)e^{-ikr_*}.
\end{equation}
at the event horizon. Here,
\begin{equation}
r_*\equiv r+2M\ln(r/2M-1),
\end{equation}
is the tortoise coordinate, reflecting the long-range nature of gravity. The incident coefficients $c_{\ell m}^{(\pm)}(k)$ are fixed by the boundary conditions, while the phase shifts $\delta_{\ell}^{(\pm)}$ and transmission coefficients $a_{\ell m}^{(\pm)}(k)$ are integration constants, determined by numerically solving the radial equations (\ref{eq_2:radial_equations}). The parity-odd and parity-even phase shifts are generally unequal, reflecting the asymmetric responses of the background spacetime to odd and even-parity perturbations.

\section{\label{sec:planar_waves}Plane gravitational waves}
As discussed above, the incident coefficients $c_{\ell m}^{(\pm)}(k)$ appearing in the asymptotic solution (\ref{eq_2:boundary_conditions_inf}) are fixed by the boundary conditions. Equivalently, in the limit where the BH mass vanishes, the asymptotic behavior (\ref{eq_2:boundary_conditions_inf}) must reduce to that of a freely propagating GW in flat spacetime.

In this work, we focus on the physically relevant case in which the GW source is sufficiently far from the lens that the incident wave can be well approximated by a plane GW. In close analogy with Eqs.\,(\ref{eq_2:Fourier}) and (\ref{eq_2:partial_wave_series}), such a plane wave may be written as
\begin{equation}
\label{eq_3:incident_Fourier}
h_{\mu\nu}^{\rm(0)}(t,\bm{r})
=\frac{1}{2\pi}\int_{-\infty}^{\infty}\tilde{h}^{(0)}_{\mu\nu}(k,\bm{r})e^{-ikt}\dd k.
\end{equation}
\begin{equation}
\label{eq_3:incident_partial_wave_series}
\begin{aligned}
\tilde{h}^{(0)}_{\mu\nu}(k,\bm{r})&=\tilde{\mathcal{A}}_{\mu\nu}(k)e^{ikr\cos\theta}\\
&=\sum_{a\ell m}
\mathscr{C}^{(a)}_{\ell m}(k,r)
\left[\mathbf{T}_{\ell m}^{(a)}(\theta,\varphi)\right]_{\mu\nu}
\end{aligned},
\end{equation}
where, without the loss of generality, the propagation direction has been chosen to lie along the positive $z$-axis. General relativity admits only two independent components in the amplitude tensor $\tilde{\mathcal{A}}_{\mu\nu}(k)$. In the transverse-traceless gauge, these are conventionally taken to be
\begin{equation}
\tilde{\mathcal{A}}_{xx}
=-\tilde{\mathcal{A}}_{yy}
\equiv\tilde{\mathcal{A}}_{+}
\end{equation}
and
\begin{equation}
\tilde{\mathcal{A}}_{xy}
=\tilde{\mathcal{A}}_{yx}
\equiv\tilde{\mathcal{A}}_{\times},
\end{equation}
with all remaining components vanishing. Under the plane-wave assumption, $\tilde{\mathcal{A}}_{+}$ and $\tilde{\mathcal{A}}_{\times}$ constitute the complete set of input parameters specifying the incident GW and scattering process.

The decomposition coefficients $\mathscr{C}^{(a)}_{\ell m}(k,r)$ appearing in Eq.\,(\ref{eq_3:incident_partial_wave_series}) are calculated via
\begin{equation}
\label{eq_3:expansion_coefficient_calculation}
\mathscr{C}^{(a)}_{\ell m}(k,r)
=\epsilon_{(a)}^{-1}
\int\eta^{\mu\alpha}\eta^{\nu\beta}
\tilde{h}_{\mu\nu}^{(0)}(t,\bm{r})
\left\{\left[\mathbf{T}^{(a)}_{\ell m}\right]_{\alpha\beta}\right\}^*\dd\Omega,
\end{equation}
where $\eta^{\mu\nu}$ denotes the Minkowski metric, $\epsilon_{(a)}$ are normalization constants defined in Eq.\,(\ref{eq_A:normalization_constants}), the asterisk indicates complex conjugation, and $\dd\Omega=r^2\sin\theta\dd{\theta}\dd{\varphi}$ is the volume element on a spherical shell. The resulting explicit expressions of Eq.\,(\ref{eq_3:expansion_coefficient_calculation}) are
\begin{widetext}
\begin{subequations}
\label{eq_3:radial_planar_GW_TT}
\begin{align}
&\mathscr{C}^{(L0)}_{\ell m}
=-\tilde{\mathcal{A}}_{\ell m}^{(+)}(k)[\sigma_{\ell}/(kr)^2]j_{\ell}(kr),
\quad(\ell\geqslant0),\\
&\mathscr{C}^{(T0)}_{\ell m}
=\tilde{\mathcal{A}}_{\ell m}^{(+)}(k)[\sigma_{\ell}/(2k^2)]j_{\ell}(kr),
\quad(\ell\geqslant0),\\
&\mathscr{C}^{(E1)}_{\ell m}
=\tilde{\mathcal{A}}_{\ell m}^{(+)}(k)[(\ell-1)(\ell+2)/k]\left[j_{\ell+1}(kr)
-(\ell+1)/(kr)j_{\ell}(kr)\right]
,\quad(\ell\geqslant1),\\
&\mathscr{C}^{(B1)}_{\ell m}
=\tilde{\mathcal{A}}_{\ell m}^{(-)}(k)
[(\ell-1)(\ell+2)/k]j_{\ell}(kr)
,\quad(\ell\geqslant1),\\
&\mathscr{C}^{(E2)}_{\ell m}
=\tilde{\mathcal{A}}_{\ell m}^{(+)}(k)r^2\left\{[1
-(\ell+1)(\ell+2)/(2(kr)^2)]j_{\ell}(kr)
+(1/kr)j_{\ell+1}(kr)\right\},
\quad(\ell\geqslant2),\\
&\mathscr{C}^{(B2)}_{\ell m}
=-\tilde{\mathcal{A}}_{\ell m}^{(-)}(k)r^2[(\ell+2)/(kr)j_{\ell}(kr)-j_{\ell+1}(kr)]
,\quad(\ell\geqslant2),
\end{align}
\end{subequations}
\end{widetext}
where 
\begin{equation}
\sigma_{\ell}\equiv(\ell-1)\ell(\ell+1)(\ell+2),
\end{equation}
\begin{equation}
\tilde{\mathcal{A}}_{\ell m}^{(\pm)}\equiv i^{\ell}\sqrt{2\pi(2\ell+1)/\sigma_{\ell}}(\tilde{\mathcal{A}}_{L}\delta_{m,-2}
\pm\tilde{\mathcal{A}}_{R}\delta_{m2}),
\end{equation}
\begin{equation}
\tilde{\mathcal{A}}_{L,R}\equiv(1/\sqrt{2})(\tilde{\mathcal{A}}_{+}\pm i\tilde{\mathcal{A}}_{\times}),
\end{equation}
and $j_{\ell}(z)$ is the spherical Bessel function of the first kind \cite{NIST:DLMF}.

To impose the boundary conditions for the scattering problem, the gauge of the incident plane GW in Eq.\,(\ref{eq_3:incident_partial_wave_series}) must be matched to the RW gauge adopted for the BH perturbations in Eq.\,(\ref{eq_2:partial_wave_series}). Applying the gauge transformation rules, summarized in Appendix \ref{app:gauge_transformation} in the flat-spacetime limit, where $f(r)=1, f'(r)=0$, the coefficients in Eq.\,(\ref{eq_3:radial_planar_GW_TT}) are mapped to their RW-gauge counterparts $\mathscr{D}_{\ell m}^{(a)}(k,r)$:
\begin{widetext}
\begin{subequations}
\label{eq_3:radial_planar_GW_RW}
\begin{align}
&\mathscr{D}_{\ell m}^{(tt)}
=-\tilde{\mathcal{A}}^{(+)}_{\ell m}(k)\{[2(kr)^2-(\ell+1)(\ell+2)]j_{\ell}(kr)+2(kr)j_{\ell+1}(kr)\},\\
&\mathscr{D}_{\ell m}^{(Rt)}
=2ikr\tilde{\mathcal{A}}^{(+)}_{\ell m}(k)
\{-(\ell+2)j_{\ell}(kr)
+(kr)j_{\ell+1}(kr)\},\\
&\mathscr{D}_{\ell m}^{(L0)}
=\tilde{\mathcal{A}}^{(+)}_{\ell m}(k)\{[(\ell+1)(\ell+2)-2(kr)^2]j_{\ell}(kr)
-2(kr)j_{\ell+1}(kr)\},\\
&\mathscr{D}_{\ell m}^{(T0)}
=r^2\tilde{\mathcal{A}}^{(+)}_{\ell m}(k)\{(\ell+1)(\ell+2)j_{\ell}(kr)
-2(kr)j_{\ell+1}(kr)\},\\
&\mathscr{D}_{\ell m}^{(B1)}
=kr^2\tilde{\mathcal{A}}^{(-)}_{\ell m}(k)j_{\ell}(kr),\\
&\mathscr{D}^{(Bt)}_{\ell m}
=ir\tilde{\mathcal{A}}^{(-)}_{\ell m}(k)[(\ell+2)j_{\ell}(kr)
-(kr)j_{\ell+1}(kr)],
\end{align}
\end{subequations}
\end{widetext}
with $\ell\geqslant2$. Defined by Eqs.\,(\ref{eq:RW_function}) and (\ref{eq:Zerilli_function}), the radial master functions for a plane GW on the flat spacetime are
\begin{subequations}
\label{eq_3:planar_GW_master}
\begin{align}
\mathscr{D}^{(-)}_{\ell m}
&=-\frac{1}{r}\mathscr{D}^{(B1)}_{\ell m}(r)=-kr\tilde{\mathcal{A}}^{(-)}_{\ell m}(k)j_{\ell}(kr),\\
\mathscr{D}_{\ell m}^{(+)}
&=\frac{1}{\lambda}\left[\frac{1}{r}
\mathscr{D}_{\ell m}^{(T0)}
+\frac{1}{ik}
\mathscr{D}_{\ell m}^{(Rt)}\right]
=2r\tilde{\mathcal{A}}^{(+)}_{\ell m}(k)j_{\ell}(kr).
\end{align}
\end{subequations}
Using the asymptotic expansion of $j_{\ell}(z)$,
\begin{equation}
j_{\ell}(z)\rightarrow z^{-1}\sin(z-\ell\pi/2),
\end{equation}
one finds the asymptotic behavior of Eq.\,(\ref{eq_3:planar_GW_master})
\begin{subequations}
\label{eq_3:planar_GW_master_asymptotic_behavior}
\begin{align}
\mathscr{D}^{(-)}_{\ell m}
&\rightarrow(-i^{\ell+1}/{2})\tilde{\mathcal{A}}_{\ell m}^{(-)}(k)
\Big[e^{-ikr}-(-1)^{\ell}e^{ikr}\Big],\\
\mathscr{D}_{\ell m}^{(+)}
&\rightarrow(i^{\ell+1}/k)\tilde{\mathcal{A}}^{(+)}_{\ell m}(k)
\Big[e^{-ikr}-(-1)^{\ell}e^{ikr}\Big].
\end{align}
\end{subequations}
Requiring that the BH radial functions $\tilde{\psi}_{\ell m}^{(\pm)}(k,r)$ reduce to $\mathscr{D}_{\ell m}^{(\pm)}(k,r)$ in the limit $M\to0$ and matching Eqs.\,(\ref{eq_2:boundary_conditions_inf}) and (\ref{eq_3:planar_GW_master_asymptotic_behavior}), the incident coefficients are finally obtained as 
\begin{subequations}
\label{eq_3:boundary_conditions}
\begin{align}
c_{\ell m}^{(-)}&=(-i^{\ell+1}/2)
\tilde{\mathcal{A}}^{(-)}_{\ell m}(k)\\
c_{\ell m}^{(+)}&=(i^{\ell+1}/k)\tilde{\mathcal{A}}^{(+)}_{\ell m}(k)
\end{align}
\end{subequations}
This completes the specification of the boundary conditions for the scattering problem. Once the incident amplitudes $\tilde{\mathcal{A}}_{+}$ and $\tilde{\mathcal{A}}_{\times}$ are given, the full radial wave functions are uniquely determined by solving Eqs.\,(\ref{eq_2:radial_equations}) together with the boundary conditions (\ref{eq_2:boundary_conditions_inf}), either analytically or numerically.

\vspace{1cm}

\section{\label{sec:scattered_GW}Scattered gravitational waves}
\subsection{Metric reconstruction}
Once the radial functions for the given parameters are obtained, the full GW metric is required to be reconstructed to extract the GW polarizations. 
The procedure of recovering the GW metric from the master variables is commonly referred to as metric reconstruction.
In the RW gauge, a detailed derivation of the reconstruction formulas is given in Ref.\,\cite{Jhingan_2003}. Here, we summarize the result as
\begin{equation}
\tilde{h}_{\ell m}^{(a)}(k,r)
=\widehat{\mathcal{J}}^{(a)}_{\ell}(k,r)\tilde{\psi}^{(\pm)}_{\ell m},
\end{equation}
where $\widehat{\mathcal{J}}^{(a)}_{\ell}(k,r)$ denote the reconstruction operators,
\begin{widetext}
\begin{subequations}
\begin{align}
&\widehat{\mathcal{J}}^{(B1)}_{\ell}(k,r)=-r/f(r),\\
&\widehat{\mathcal{J}}^{(Bt)}_{\ell}(k,r)=[f(r)/(ik)](1+r\partial_{r}),\\
&\widehat{\mathcal{J}}^{(T0)}_{\ell}(k,r)=r\{(1/\Lambda)[\sigma_{\ell}/4+3\lambda(M/r)+6(M/r)^2]+rf(r)\partial_{r}\},\\
&\widehat{\mathcal{J}}^{(Rt)}_{\ell}(k,r)=-ik\left\{[1/\Lambda f(r)][\lambda-3\lambda(M/r)-3(M/r)^2]+r\partial_{r}\right\},\\
&\widehat{\mathcal{J}}^{(L0)}_{\ell}(k,r)
=-[r/f^2(r)]
[k^2-(1/2)V_{\ell}^{(+)}(r)]-[1/f(r)][M/r-\lambda f(r)/\Lambda]\partial_{r},\\
&\widehat{\mathcal{J}}^{(tt)}_{\ell}(k,r)=f^2(r)\widehat{\mathcal{J}}^{(+,L0)}_{\ell}(k,r).
\end{align}
\end{subequations}
\end{widetext}
It should be emphasized that the GW metric components $\tilde{h}_{\ell m}^{(a)}(k,r)$ are gauge dependent and therefore do not directly correspond to observable GW effects. To extract physically meaningful quantities, we instead compute the Weyl scalars.

\subsection{Weyl scalars}
The Weyl scalars are defined as \cite{Chandrasekhar1983}
\begin{subequations}
\label{eq_4:Weyl_definitions}
\begin{align}
\psi_0&\equiv-C^{\rm(GW)}_{\alpha\beta\gamma\delta}z_{(1)}^{\alpha}z_{(3)}^{\beta}z_{(1)}^{\gamma}z_{(3)}^{\delta},\\
\psi_1&\equiv-C^{\rm(GW)}_{\alpha\beta\gamma\delta}z_{(1)}^{\alpha}z_{(2)}^{\beta}z_{(1)}^{\gamma}z_{(3)}^{\delta},\\
\psi_2&\equiv-C^{\rm(GW)}_{\alpha\beta\gamma\delta}z_{(1)}^{\alpha}z_{(3)}^{\beta}z_{(2)}^{\gamma}z_{(4)}^{\delta},\\
\psi_3&\equiv-C^{\rm(GW)}_{\alpha\beta\gamma\delta}z_{(1)}^{\alpha}z_{(2)}^{\beta}z_{(2)}^{\gamma}z_{(4)}^{\delta},\\
\psi_4&\equiv-C^{\rm(GW)}_{\alpha\beta\gamma\delta}z_{(2)}^{\alpha}z_{(4)}^{\beta}z_{(2)}^{\gamma}z_{(4)}^{\delta},
\end{align}
\end{subequations}
where $z_{(a)}^{\alpha}$ is the Newman-Penrose (NP) null tetrad and $C^{\rm(GW)}_{\alpha\beta\gamma\delta}$ is the Weyl tensor of the linear perturbation, which coincides with the Riemann tensor $R^{\rm(GW)}_{\alpha\beta\gamma\delta}$ in vacuum. In a flat background, all five Weyl scalars are gauge invariant, while in a curved spacetime, only $\psi_4$ is strictly gauge invariant. Nevertheless, for observers located sufficiently far from the BH, the Weyl scalars are approximately gauge invariant and provide a useful description of GW polarizations.

Compatible with the Schwarzschild background, we first adopt the Kinnersley tetrad \cite{Kinnersley_1969}, $z_{(a)}^{\alpha}=e_{(a)}^{\alpha}=\{\bm{l},\bm{n},\bm{m},\bm{m}^*\}$ with
\begin{subequations}
\begin{align}
\bm{l}&=\{f^{-1}(r),1,0,0\}\\
\bm{n}&=\frac{1}{2}\{1,-f(r),0,0\}\\
\bm{m}&=\frac{1}{\sqrt{2}r}\{0,0,1,i\csc\theta\},
\end{align}
\end{subequations}
to calculate the Weyl scalars. For notational convenience, we denote them by $\Psi_n$ $(n=0,1,2,3,4)$.
Substituting the expansion of the GW metric, (\ref{eq_2:Fourier}) and (\ref{eq_2:partial_wave_series}), into the linearized Riemann tensor, projecting onto the NP tetrad $e_{(a)}^{\alpha}$ via Eq.\,(\ref{eq_4:Weyl_definitions}), and applying metric reconstruction, the Weyl scalars are expressed in terms of the master variables as
\begin{equation}
\label{eq_4:Weyl_Fourier}
\Psi_{n}(t,\bm{r})=\frac{1}{2\pi}\int_{-\infty}^{\infty}\tilde{\Psi}_n(k,\bm{r})e^{-ikr}\dd{k},
\end{equation}
\begin{equation}
\label{eq_4:Weyl_partial_wave_series}
\tilde{\Psi}_{n}(k,\bm{r})=\sum_{\ell=2}^{\infty}\sum_{m=-\ell}^{\ell}\tilde{\Psi}_{n,\ell m}(k,\bm{r}),
\end{equation}
where the frequency-domain components are 
\begin{widetext}
\begin{subequations}
\label{eq_4:Weyl_Fourier}
\begin{align}
\tilde{\Psi}_{4,\ell m}&=\sqrt{\sigma_{\ell}}\left[\mathcal{Z}^{(-)}_{4,\ell m}(k,r)+\mathcal{Z}^{(+)}_{4,\ell m}(k,r)\right]{_{-2}Y_{\ell m}}(\theta,\varphi),\\
\tilde{\Psi}_{3,\ell m}&=\sqrt{2\ell(\ell+1)}\left[\mathcal{Z}^{(-)}_{3,\ell m}(k,r)+\mathcal{Z}^{(+)}_{3,\ell m}(k,r)\right]{_{-1}Y_{\ell m}}(\theta,\varphi),\\
\tilde{\Psi}_{2,\ell m}&=\left[\mathcal{Z}^{(-)}_{2,\ell m}(k,r)+\mathcal{Z}^{(+)}_{2,\ell m}(k,r)\right]{Y_{\ell m}}(\theta,\varphi),\\
\tilde{\Psi}_{1,\ell m}&=\sqrt{2\ell(\ell+1)}\left[\mathcal{Z}^{(-)}_{1,\ell m}(k,r)+\mathcal{Z}^{(+)}_{1,\ell m}(k,r)\right]{_{+1}Y_{\ell m}}(\theta,\varphi),\\
\tilde{\Psi}_{0,\ell m}&=\sqrt{\sigma_{\ell}}\left[\mathcal{Z}^{(-)}_{0,\ell m}(k,r)+\mathcal{Z}^{(+)}_{0,\ell m}(k,r)\right]{_{+2}Y_{\ell m}}(\theta,\varphi),
\end{align}
\end{subequations}
where the angular dependence is given by ${_sY_{\ell m}}(\theta,\varphi)$, known as the spin-weighted spherical harmonics, and
\begin{subequations}
\begin{align}
&16r^3\times\mathcal{Z}_{4,\ell m}^{(-)}(k,r)
=(2/k)[r^2V^{(-)}_{\ell}+2ikr(1-3M/r)-2(kr)^2]\tilde{\psi}_{\ell m}^{(-)}
+(4/k)f(r)(1-3M/r+ikr)r\partial_{r}\tilde{\psi}_{\ell m}^{(-)},\\
&\begin{aligned}
16r^3\times\mathcal{Z}_{4,\ell m}^{(+)}(k,r)&=\{r^2V^{(+)}_{\ell}+2ikr/\Lambda[\lambda-3\lambda(M/r)-3(M/r)^2]-2(kr)^2\}\tilde{\psi}_{\ell m}^{(+)}\\
&\qquad+2f(r)\left\{(1/\Lambda)[\lambda-3\lambda(M/r)-3(M/r)^2]+ikr\right\}r\partial_{r}\tilde{\psi}_{\ell m}^{(+)},
\end{aligned}\\
&8r^3\times\mathcal{Z}^{(-)}_{3,\ell m}(k,r)
=(2/k)[f(r)(M/r)+ikr(\lambda+M/r)]\tilde{\psi}_{\ell m}^{(-)}
+(2/k)f(r)(\lambda+M/r)r\partial_{r}\tilde{\psi}_{\ell m}^{(-)},\\
&8r^3\times\mathcal{Z}^{(+)}_{3,\ell m}(k,r)=[-3f(r)(M/r)+ikr\Lambda]\tilde{\psi}_{\ell m}^{(+)}+f(r)\Lambda r\partial_{r}\tilde{\psi}_{\ell m}^{(+)},\\
&16r^3\times\mathcal{Z}^{(-)}_{2,\ell m}(k,r)
=(4/k)\sigma_{\ell}\tilde{\psi}_{\ell m}^{(-)},\\
&\begin{aligned}
16r^3\times\mathcal{Z}^{(+)}_{2,\ell m}(k,r)
&=(1/\Lambda)[4\ell(\ell+1)\lambda^2+10\sigma_{\ell}(M/r)+24(\ell^2+\ell+1)(M/r)^2-48(M/r)^3]\tilde{\psi}_{\ell m}^{(+)}\\
&\qquad-8f(r)(M/r)r\partial_{r}\tilde{\psi}_{\ell m}^{(+)},
\end{aligned}\\
&\mathcal{Z}^{(\pm)}_{1,\ell m}(k,r)=[2/f(r)][\mathcal{Z}^{(\pm)}_{3,\ell m}(k,r)]^*,\\
&\mathcal{Z}^{(\pm)}_{0,\ell m}(k,r)=[2/f(r)]^2[\mathcal{Z}^{(\pm)}_{4,\ell m}(k,r)]^*.
\end{align}
\end{subequations}
\end{widetext}

For the scattering processes considered here, the components propagating along the initial propagation direction ($z$ axis) dominate the scattered GWs. To define the GW polarization in a convenient manner, we introduce an alternative tetrad $z_{(a)}^{\alpha}=\hat{e}_{(a)}^{\alpha}=\{\hat{\bm{l}},\hat{\bm{n}},\hat{\bm{m}},\hat{\bm{m}}^*\}$, with
\begin{subequations}
\label{eq_4:tetrad_z}
\begin{align}
\hat{\bm{l}}&=\frac{1}{\sqrt{2}}(1,0,0,1)\\
\hat{\bm{n}}&=\frac{1}{\sqrt{2}}(1,0,0,-1)\\
\hat{\bm{m}}&=\frac{1}{\sqrt{2}}(0,1,i,0),   
\end{align}
\end{subequations}
in which the first null leg is aligned with the $z$ axis.

The NP components of the linearized Riemann tensor transform between the tetrads $e^{\alpha}_{(a)}$ and $\hat{e}^{\alpha}_{(a)}$ according to
\begin{equation}
\hat{R}^{\rm(GW)}_{(a)(b)(c)(d)}=\Lambda^{(i)(j)(k)(m)}_{(a)(b)(c)(d)}R^{\rm(GW)}_{(i)(j)(k)(m)},
\end{equation}
with
\begin{equation}
\begin{aligned}
\Lambda^{(i)(j)(k)(m)}_{(a)(b)(c)(d)}&\equiv\left(\frac{\partial \hat{x}^{\alpha}}{\partial x^{\mu}}
\frac{\partial \hat{x}^{\beta}}{\partial x^{\nu}}
\frac{\partial \hat{x}^{\gamma}}{\partial x^{\rho}}
\frac{\partial \hat{x}^{\delta}}{\partial x^{\sigma}}\right)\\
&\times\left[\hat{e}^{\,\mu}_{(a)}\hat{e}^{\,\nu}_{(b)}\hat{e}^{\,\rho}_{(c)}\hat{e}^{\,\sigma}_{(d)}\right]
\left[e^{(i)}_{\alpha}e^{(j)}_{\beta}e^{(k)}_{\gamma}e^{(m)}_{\delta}\right],
\end{aligned}
\end{equation}
where $\partial \hat{x}^{\alpha}/\partial x^{\mu}$ denotes the Jacobian between the Schwarzschild and Cartesian coordinates. Accordingly, the Weyl scalars transform as
\begin{equation}
\label{eq_4:Weyl_transformation}
\hat{\Psi}_{2-m}=\Lambda_{2ms}\Psi_{2-s},
\end{equation}
with
\begin{equation}
\Lambda_{\ell ms}=(-1)^{s+m}2^{-s/2}\sqrt{\frac{(\ell+m)! (\ell-m)!}{(\ell+s)!(\ell-s)!}}D^{\ell}_{ms}(\varphi,\theta,0),
\end{equation}
where $s,m=\{-2,-1,0,1,2\}$ and $D^{j}_{mm'}$ is the Wigner-D function \cite{NIST:DLMF}.

\subsection{GW polarization}
The observable GW polarizations are extracted from the geodesic deviation equation,
\begin{equation}
\ddot{\zeta}^j+\hat{R}^{\rm(GW)}_{0j0k}\zeta^{k}=0,
\end{equation}
where $\zeta^j$ denotes the deviation vector and the overdot represents the time derivative. In terms of the electric components of the Riemann tensor, the GW polarizations are defined as \cite{Hyun_2019}
\begin{subequations}
\label{eq_4:GW_polarizations}
\begin{align}
\ddot{h}_{+}&=\mathrm{Re}\hat{\Psi}_{4}+\mathrm{Re}\hat{\Psi}_{0},\label{eq_4:h_plus}\\
\ddot{h}_{\times}&=-(\mathrm{Im}\hat{\Psi}_{4}-\mathrm{Im}\hat{\Psi}_{0}),\label{eq_4:h_cross}\\
\ddot{h}_{x}&=(1/2)(\mathrm{Re}\hat{\Psi}_{1}
+\mathrm{Re}\hat{\Psi}_{3}),\\
\ddot{h}_{y}&=(1/2)(\mathrm{Im}\hat{\Psi}_{1}
-\mathrm{Im}\hat{\Psi}_{3}),\\
\ddot{h}_{b}&=(1/2)\mathrm{Re}\hat{\Psi}_{2},\\
\ddot{h}_{L}&=\mathrm{Re}\hat{\Psi}_{2}.
\end{align}
\end{subequations}
It should be noted that the propagation direction of scattered GWs is not uniquely defined, as the wavefront is distorted and smeared by the scattering process, as shown in the scalar scattering \cite{Li_2025_b}. Here, we adopt a natural and convenient prescription in which the first null leg $\hat{\bm{l}}$ is aligned with the null direction of the incident wave at future infinity. An inevitable consequence of this choice is the emergence of richer oscillatory structures beyond the conventional $+$ and $\times$ modes. These features are purely environmental effects and do not represent additional physical degrees of freedom. In the present work, we focus exclusively on the $+$ and $\times$ modes and compare them with the Kirchhoff integral \cite{Takahashi_2003} and other conventional approaches \cite{Pijnenburg_2024b,Chan_2025}, setting aside the complexities associated with apparent polarization modes.

\section{\label{sec:numerical_results}Numerical Results}

\begin{figure*}[htbp]
    \centering
    \includegraphics[width=1.0\linewidth]{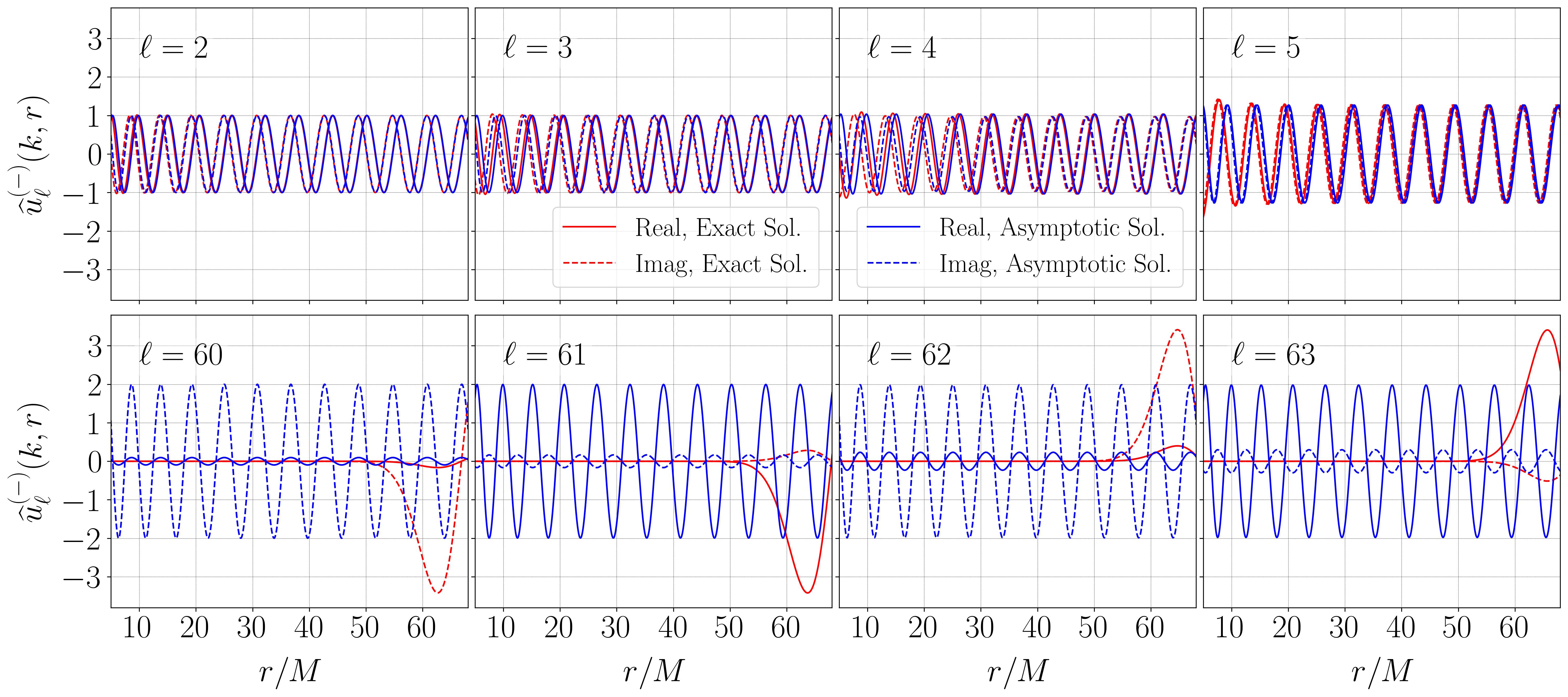}
    \caption{Validation of the asymptotic expansion for the normalized radial function $\widehat{u}_{\ell}^{(+)}(k,r)$, with $k=1.0/M$.}
    \label{fig:validation_asymptotic_expansion}
\end{figure*}

\begin{figure*}[htbp]
    \centering
    \includegraphics[width=1.0\linewidth]{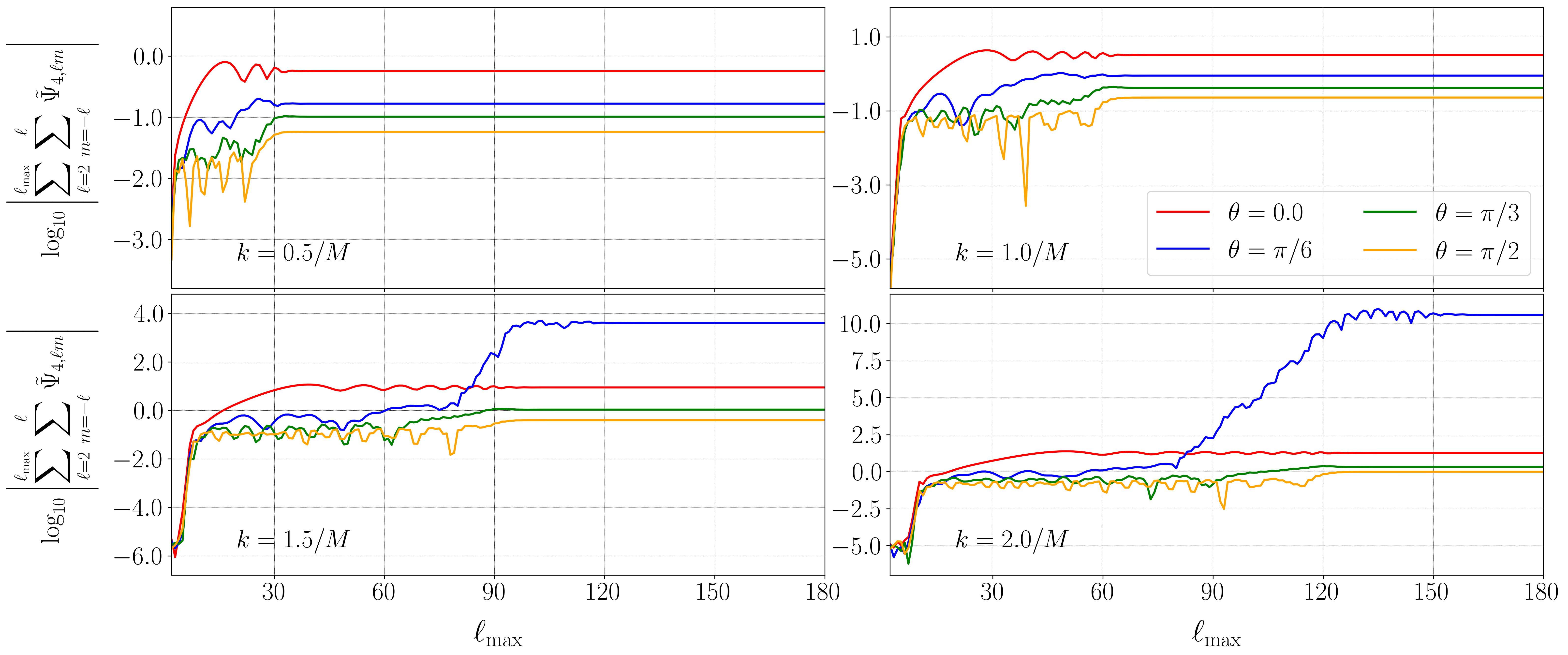}
    \caption{Convergence of the partial-wave series for $\tilde{\Psi}_4$ at selected frequencies $k$ and scattering angles $\theta$. The observer is located at $r=60.0M$.}
    \label{fig:convergence}
\end{figure*}

\subsection{\label{subsec:convergence}The convergence of partial-wave series}

We first summarize the computational procedure for scattered waves. The GW polarization modes are defined in Eq.\,(\ref{eq_4:GW_polarizations}), where the Weyl scalars are calculated from Eq.\,(\ref{eq_4:Weyl_Fourier}) and transformed through Eq.\,(\ref{eq_4:Weyl_transformation}). To achieve this, we numerically solve the RW and Zerilli equations (\ref{eq_2:radial_equations}) to find radial wave function $\tilde{\psi}_{\ell m}^{(\pm)}$, with their outer boundary conditions being (\ref{eq_2:boundary_conditions_inf}) and the inner ones being (\ref{eq_2:boundary_conditions_horizon}), where the incident coefficients $c_{\ell m}^{(\pm)}$ are determined by the plane-wave assumption, by Eq.\,(\ref{eq_3:boundary_conditions}). 

This work computes the scattered GWs for frequencies $k$ ranging from $0.1/M$ to $4.0/M$ and multipoles $\ell$ up to $\ell_{\max}$. The choice of $\ell_{\max}$ ensures convergence of the partial-wave series in Eqs.\,(\ref{eq_2:partial_wave_series}) and (\ref{eq_4:Weyl_partial_wave_series}). Empirically, one finds $\ell_{\max}\sim kr$, where $r$ denotes the radius of the observer \cite{Li_2025_b,Kubota_2024_b}. In our calculations, the free parameters are arbitrarily set to $\tilde{\mathcal{A}}_{+}=0.9+1.1i$ and $\tilde{\mathcal{A}}_{\times}=0.4+0.6i$ for numerical demonstration, without loss of generality of our conclusions.

It is critical to distinguish our approach from previous studies \cite{Pijnenburg_2024b,Chan_2025,Saketh_2025}, which assumed that the observer is asymptotically far from the lens and employed the asymptotic expansion of the radial functions $u_{\ell}^{(\pm)}$, as given in Eq.\,(\ref{eq_2:boundary_conditions_inf}), when resuming the partial-wave series (\ref{eq_2:partial_wave_series}). 
As demonstrated in our previous work \cite{Li_2025_b}, such an asymptotic expansion is valid only in the regime $kr \gg \ell$. Applying it to high-$\ell$ modes (i.e., $\ell\sim\ell_{\max}$) substantially overestimates their contributions, causing divergences in the partial-wave series and in the Poisson spot.

Figure\,\ref{fig:validation_asymptotic_expansion} compares the exact numerical solution of the normalized radial function,
\begin{equation}
\widehat{u}_{\ell}^{(-)}\equiv\left[c_{\ell m}^{(-)}\right]^{-1}\tilde{\psi}_{\ell m}^{(-)},
\end{equation}
with its asymptotic behavior
\begin{equation}
\label{eq_5:boundary_conditions_inf}
\widehat{u}_{\ell}^{(-)}(k,r\rightarrow\infty)\rightarrow e^{-ikr_*}-(-1)^{\ell}e^{2i\delta^{(-)}_{\ell}}e^{ikr_*}.
\end{equation}
At $kr\sim 60$, the asymptotic expansion accurately reproduces the low-$\ell$ modes ($\ell \ll 60$), but deviates significantly for high-$\ell$ modes ($\ell \gtrsim 60$).

Alternatively, to avoid these unphysical divergences, we instead place the observer at a finite radius and compute the partial-wave series in Eqs.\,(\ref{eq_2:partial_wave_series}) and (\ref{eq_4:Weyl_partial_wave_series}) without invoking the asymptotic expansion. The convergence of $\tilde{\Psi}_4$ under this scheme is illustrated in Fig.\,\ref{fig:convergence} for various frequencies and scattering angles, with the observer located at $r=60.0M$.

\begin{figure*}[htbp]
    \centering
    \includegraphics[width=1.0\linewidth]{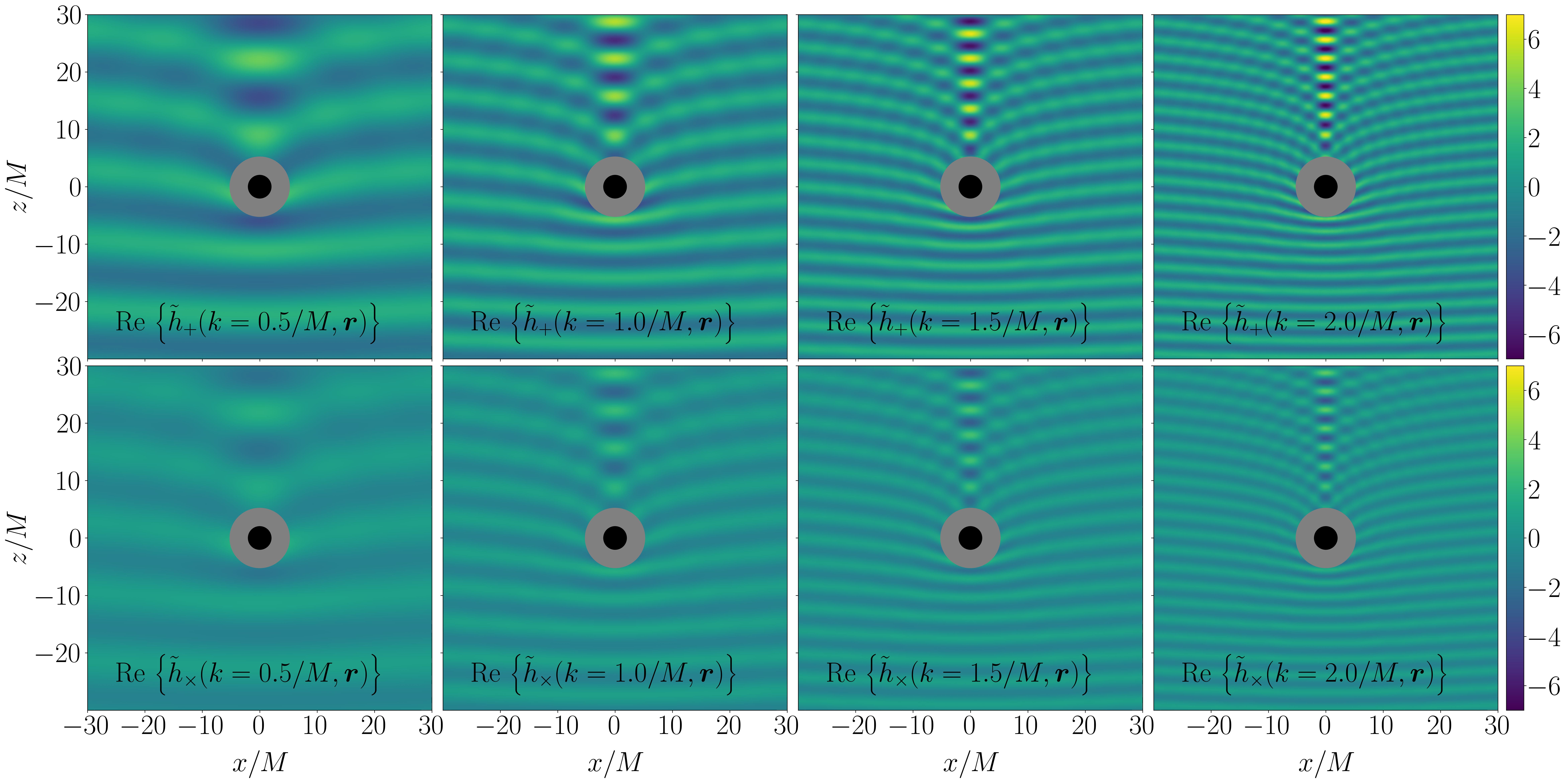}
    \caption{Wave fields of the $+$ and $\times$ modes for scattered GWs. The black and gray regions represent the BH event horizon and light ring, respectively.}
    \label{fig:wave_field}
\end{figure*}

\begin{figure*}[htbp]
    \centering
    \includegraphics[width=1.0\linewidth]{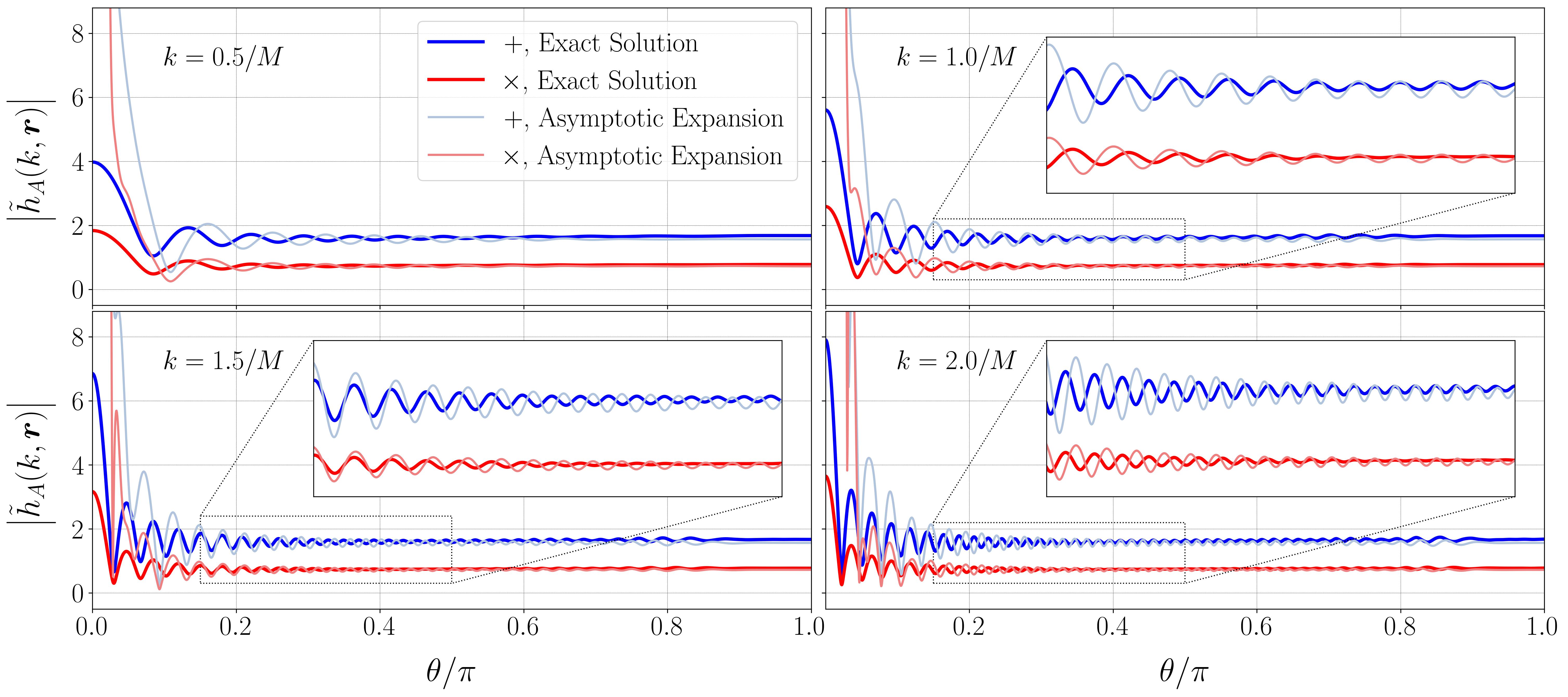}
    \caption{Diffraction patterns of the $+$ and $\times$ modes for scattered GWs, compared with results obtained using the conventional asymptotic approach (see Appendix \ref{app:conventional_computations}).}
    \label{fig:pattern}
\end{figure*}

\subsection{Wave field}
We first present the wave fields of the $+$ and $\times$ polarizations of the scattered GWs in Fig.\,\ref{fig:wave_field}, for various frequencies, within a region of size $60M\times60M$. The main features of the spatial distribution are summarized as follows. (1) A bright Poisson spot forms near the optical axis as a result of signal focusing during the scattering process. (2) The wavefront is significantly distorted and smeared by the reflected spherical waves centered on the BH, producing a network-like structure in the far-axis region corresponding to large scattering angles.

Another important observation is that the propagation direction of the scattered GW is no longer defined unambiguously in a global or intuitive sense. Instead, it depends on the observer’s spacetime location. This leads to the conceptual difficulty encountered in Sec.\,\ref{sec:scattered_GW}: how should one construct an NP tetrad to define the GW polarization? The scheme adopted here is a natural and convenient choice, based on the propagation direction of the incident wave, which remains dominant even after scattering. The tetrad used in this work is given in Eq.\,(\ref{eq_4:tetrad_z}), and the corresponding GW polarizations are defined in Eq.\,(\ref{eq_4:GW_polarizations}).
As a consequence, the Weyl scalars $\Psi_n$ $(n=0,1,2,3)$ do not vanish, leading to non-zero apparent vector and scalar polarizations that are unphysical. The wave fields associated with these apparent modes are presented in the Appendix \ref{app:wave_field}. Although such apparent polarizations play a role, a detailed examination of them is beyond the scope of this paper, which aims to elucidate computations of $+$ and $\times$ modes.

\subsection{The diffraction pattern}
The diffraction pattern characterizes the angular distribution of GW amplitudes. Figure \ref{fig:pattern} displays the $+$ and $\times$ modes at $r=60.0M$ for various frequencies, together with a comparison to results obtained using the conventional computational approach for BH scattering, reviewed in Appendix \ref{app:conventional_computations}.

As discussed above, the asymptotic expansion employed in the conventional approach leads to divergences when resumming the partial-wave series, particularly at the Poisson spot. To eliminate these singular behaviors, a variety of regularization techniques have been proposed, including series reduction \cite{Yennie_1954,Stratton_2020,Dolan_2008b}, Ces\`{a}ro summation \cite{Pijnenburg_2024b}, the complex angular momentum method \cite{Andersson_1994,Folacci_2019a,Folacci_2019b}, and the Fresnel half-wave-zone method \cite{Zhang_Fan_2021}. The diffraction patterns obtained using the asymptotic expansion and shown in Fig.\,\ref{fig:pattern} are based on the second-order series reduction. In the far-axis region, corresponding to relatively large scattering angles (e.g., $\theta\gtrsim30\text{\textdegree}$), the asymptotic solutions reproduce the main qualitative features of the exact results. However, their validity breaks down in the near-axis region (e.g., $\theta\lesssim30\text{\textdegree}$), where they fail to provide any reasonable prediction for the behavior of the scattered waves. This pathology, commonly referred to as the Poisson-spot or forward-scattering divergence, persists even when regularization techniques are applied within the conventional framework. We have revealed that this divergence originates from the inappropriate use of the asymptotic expansion at small scattering angles, and it is therefore naturally avoided by abandoning the asymptotic expansion \cite{Li_2025_b}.

\begin{table}[ht]
\centering
\renewcommand{\arraystretch}{1.3}
\setlength{\tabcolsep}{12pt}
\begin{tabular}{cccc}
\hline\hline
region & $x\,(M)$ & $\theta\,(\text{\textdegree})$ & $\xi/\xi_0$ \\
\hline
\multirow{4}{*}{near-axis} 
& 0.00 & 0.00000 & 0.0000\\
& 1.00 & 1.90915 & 0.0913\\
& 2.00 & 3.81407 & 0.1828\\
& 3.00 & 5.71059 & 0.2745\\
\hline
\multirow{4}{*}{far-axis}
& 10.0 & 18.4349 & 0.9372\\
& 15.0 & 26.5651 & 1.4479\\
& 20.0 & 33.6901 & 2.0015\\
& 25.0 & 39.8056 & 2.6038\\
\hline\hline
\end{tabular}
\caption{Observer positions used for computing transmission factors.}
\label{tab:observer_position}
\end{table}

\begin{figure*}[htbp]
    \centering
    \includegraphics[width=1.0\linewidth]{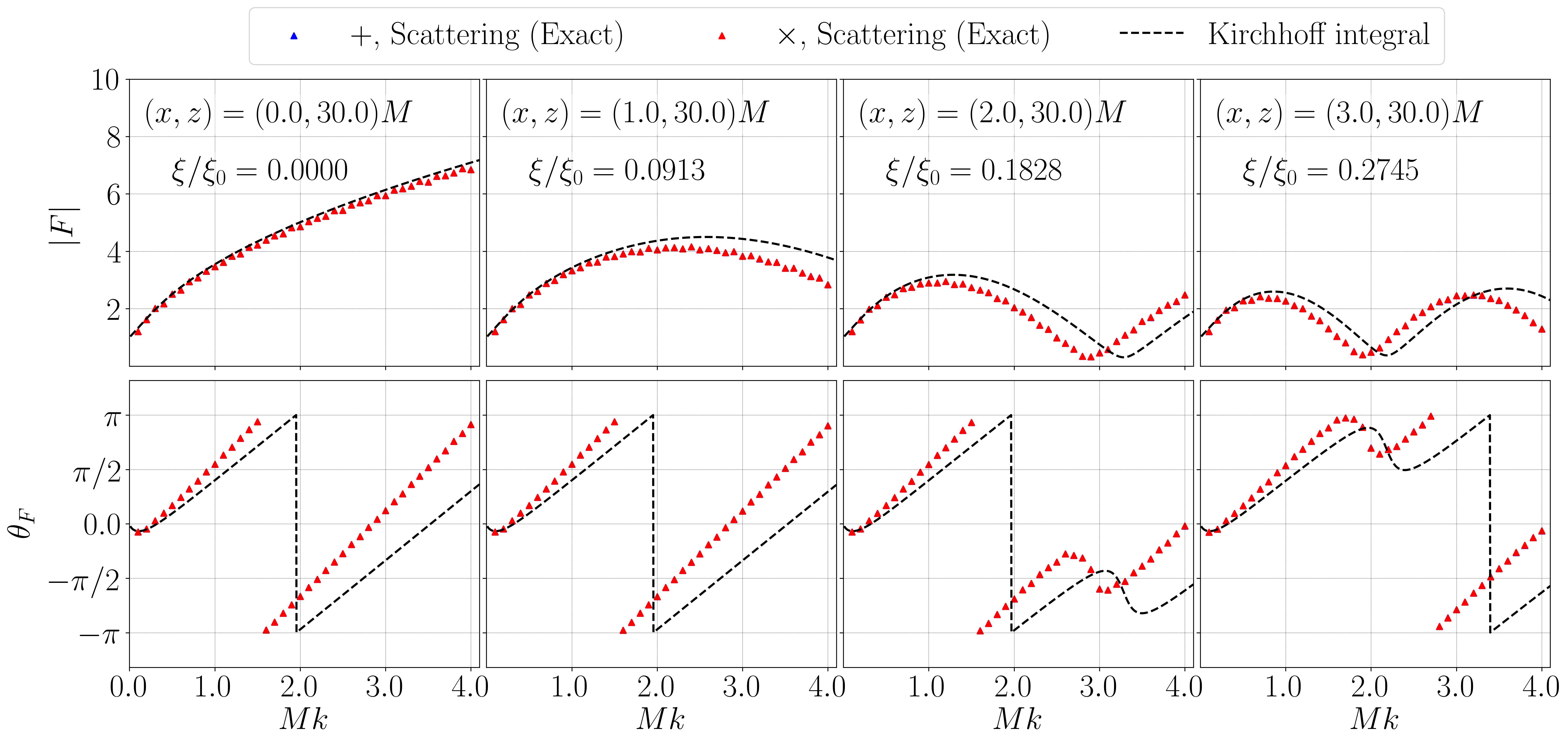}
    \caption{Comparison of transmission factors from BH scattering (dots) and Kirchhoff integral (dashed lines) in the near-axis region.}
    \label{fig:transmission_factor_near_axis}
\end{figure*}

\begin{figure*}[htbp]
    \centering
    \includegraphics[width=1.0\linewidth]{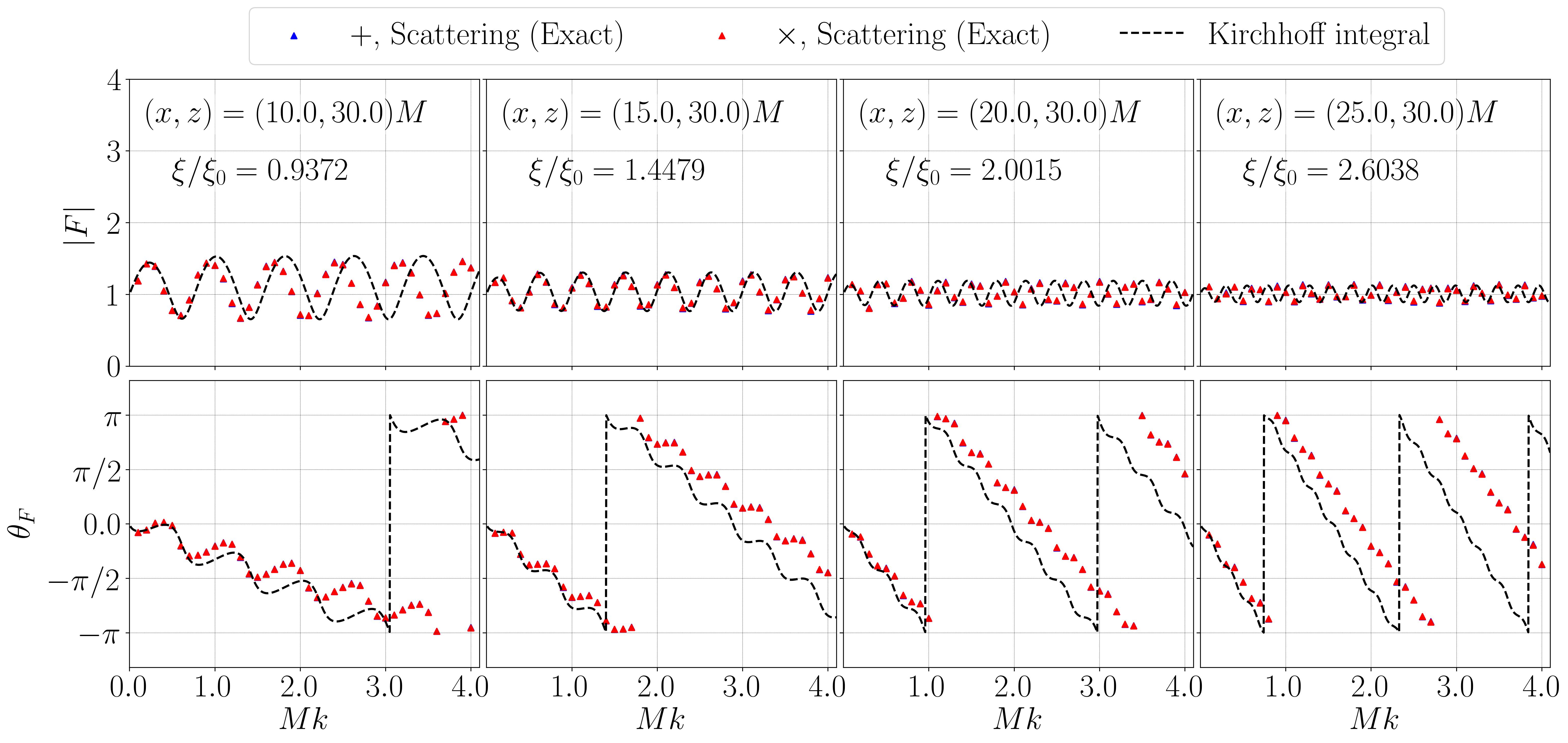}
    \caption{Comparison of transmission factors from BH scattering (dots) and Kirchhoff integral (dashed lines) in the far-axis region.}
    \label{fig:transmission_factor_far_axis}
\end{figure*}

\subsection{Transmission factor}
The transmission factor is defined as the ratio between the lensed and unlensed frequency-domain waveforms,
\begin{equation}
\tilde{h}_{+,\times}=F_{+,\times}\tilde{h}^{(0)}_{+,\times}.
\end{equation}
with
\begin{equation}
\label{eq_5:incident_polarization}
\tilde{h}^{(0)}_{+,\times}
=\tilde{\mathcal{A}}_{+,\times}e^{ikr\cos\theta},
\end{equation}
where $\tilde{h}_{+,\times}$ are the Fourier transforms of the waveforms defined in Eqs.\,(\ref{eq_4:h_plus}) and (\ref{eq_4:h_cross}).
The magnitude $|F|$ characterizes the GW signal magnification, while its argument $\theta_F$ corresponds to the phase shift.

For a point-mass lens in the weak-field limit, the transmission factor $F_{+,\times}(k)$ can also be obtained using the Kirchhoff integral approach \cite{Takahashi_2003,Meena_2019,Guo_2020} 
\begin{equation}
\label{eq:Kirchhoff}
\begin{aligned}
F_{+,\times}&=e^{\pi\gamma/2}\left(-\gamma\right)^{-i\gamma}\Gamma(1+i\gamma)\\
&\qquad\times{_1F_1}(-i\gamma,1;-i\gamma(\xi/\xi_0)^2),
\end{aligned}
\end{equation}
where $\gamma\equiv-2Mk$, $\xi/\xi_0=(1/2)(r/M)^{1/2}\tan\theta$ is the normalized angular coordinate, and ${_1F_1}(\cdots,\cdots;\cdots)$ is the Kummer hypergeometric function \cite{NIST:DLMF}. 

Figures \ref{fig:transmission_factor_near_axis} and \ref{fig:transmission_factor_far_axis} compare the transmission factors obtained from BH scattering with those derived from the Kirchhoff diffraction integral. The comparisons are divided into two groups, corresponding to the near-axis and far-axis regions, with the observer positions listed in Table \ref{tab:observer_position}. While these two groups of results, obtained from scattering formalism and Kirchhoff integral, exhibit qualitatively similar oscillatory behavior and overall trends across all considered observer positions, significant quantitative deviations are commonly observed, especially in the high-frequency regime. These discrepancies stem from intrinsic limitations of the Kirchhoff integral method, including (1) the reliance on geometric-optics approximations, (2) the neglect of the long-range nature of gravitational interactions, (3) the omission of polarization evolution associated with the spin-2 nature of GWs, and (4) the approximation of the gravitational potential by its Newtonian limit.

\section{\label{sec:summary}Summary}
GW lensing has attracted increasing attention in recent years. However, the traditional theoretical frameworks, namely geometric optics and the Kirchhoff diffraction integral, are unable to simultaneously capture the wave nature and the spin-2 character of scattered GWs. By contrast, within the alternative framework of BH scattering, which is formulated from the first principle and relies solely on the linear perturbation theory, the wave and polarization signatures of lensed GWs remain insufficiently understood, particularly in the vicinity of the optical axis \cite{Pijnenburg_2024b,Chan_2025,Saketh_2025}.

In this work, we present a more rigorous calculation of GW scattering by a Schwarzschild BH, covering both on-axis and off-axis regions. By discarding the asymptotic expansion of the radial wave functions, we avoid the divergences associated with the partial-wave series and the Poisson spot without invoking any regularization procedure. The origin of these divergences has been discussed in our previous work \cite{Li_2025_b} and is revisited in Sec.\,\ref{subsec:convergence}. The convergence of the series in our calculation are explicitly confirmed, as demonstrated in Fig.\,\ref{fig:convergence}.

Through numerical calculations, we first present the wave fields of the scattered $+$ and $\times$ polarizations, clearly exhibiting the formation of the Poisson spot and the distortion of the wavefronts (see Fig.\,\ref{fig:wave_field}). In addition to these two physical polarizations, we also extract the apparent polarization modes and display their corresponding wave fields in Appendix \ref{app:wave_field}. We then present the diffraction patterns—namely, the angular distributions of GW amplitudes—and compare them with those obtained using the conventional computational procedure in Fig.\,\ref{fig:pattern}. The latter approach is reviewed in Appendix \ref{app:conventional_computations}, where the divergences induced by the asymptotic expansion are eliminated using the second-order series reduction method \cite{Stratton_2020}, summarized in Appendix \ref{app:series_reduction_method}. This method performs well in the far-axis region but becomes ineffective in the vicinity of the optical axis \cite{Chan_2025}. As shown in Fig.\,\ref{fig:pattern}, the asymptotic solutions successfully reproduce the main oscillatory features of the scattered GWs and are broadly consistent with the exact results at large scattering angles. However, they exhibit singular behavior near the optical axis and fail to provide physically meaningful predictions in this regime. In contrast, our exact solutions remain well behaved and correctly capture the Poisson spot.

We further compare our results with those obtained from the Kirchhoff diffraction integral \cite{Takahashi_2003}, one of the most widely used methods in GW lensing modeling. To this end, we compute the transmission factors, which characterize the amplitude magnification and phase shift of GWs induced by lensing, over a range of scattering angles and frequencies. The comparison reveals non-negligible deviations between the two approaches, especially in the high-frequency regime, e.g., $k\gtrsim1.0/M$. These discrepancies originate from several inherent approximations in the Kirchhoff integral, including the adoption of geometric-optics assumptions, the neglect of polarization evolution and the long-range nature of gravity, and the treatment of the lens as a weak-gravity object.

Although further developments are required to construct accurate lensed waveform templates for future GW-lensing searches, this work establishes a solid foundation for rigorous GW-scattering calculations and provides a more comprehensive understanding of GW lensing. Future work will focus on solving the high-$\ell$ perturbation equations and extending the framework to Kerr lensing and to larger, astrophysically relevant scales.

\begin{acknowledgements}
We thank Takahiro Tanaka, Jianhua He, Xi-Long Fan, and Xian Chen for their helpful discussions.     
This work was supported by the
National Natural Science Foundation of China with grant Nos.
12325301 and 12273035.
Z. L. is supported by the China Postdoctoral Science
Foundation under Grant No. 2025M783223.
S. H. is supported by the National Key Research and Development Program of China, grant No.~2024YFC2207400.
\end{acknowledgements}

\begin{widetext}
\appendix

\section{\label{app:spherical_harmonics}Spherical harmonics}
The definitions of harmonic tensors are listed as follows,
\begin{subequations}
\begin{align}
\mathbf{T}^{(tt)}_{\ell m}
&=\left(\begin{array}{cccc}
1 & 0 & 0 & 0\\
0 & 0 & 0 & 0\\
0 & 0 & 0 & 0 \\
0 & 0 & 0 & 0 
\end{array}\right)Y_{\ell m},\\
\mathbf{T}^{(Rt)}_{\ell m}
&=\left(\begin{array}{cccc}
0 & 1 & 0 & 0\\
1 & 0 & 0 & 0\\
0 & 0 & 0 & 0 \\
0 & 0 & 0 & 0 
\end{array}\right)Y_{\ell m},\\
\mathbf{T}^{(L0)}_{\ell m}
&=\left(\begin{array}{cccc}
0 & 0 & 0 & 0\\
0 & 1 & 0 & 0\\
0 & 0 & 0 & 0 \\
0 & 0 & 0 & 0 
\end{array}\right)Y_{\ell m},\\
\mathbf{T}^{(T0)}_{\ell m}
&=\left(\begin{array}{cccc}
0 & 0 & 0 & 0\\
0 & 0 & 0 & 0\\
0 & 0 & 1 & 0 \\
0 & 0 & 0 & \sin^2\theta 
\end{array}\right)Y_{\ell m},\\
\mathbf{T}^{(Et)}_{\ell m}
&=\left(\begin{array}{cccc}
0 & 0 & \partial_{\theta} & \partial_{\varphi}\\
0 & 0 & 0 & 0\\
\partial_{\theta} & 0 & 0 & 0 \\
\partial_{\varphi} & 0 & 0 & 0 
\end{array}\right)Y_{\ell m},\\
\mathbf{T}^{(E1)}_{\ell m}
&=\left(\begin{array}{cccc}
0 & 0 & 0 & 0\\
0 & 0 & \partial_{\theta} & \partial_{\varphi}\\
0 & \partial_{\theta} & 0 & 0 \\
0 & \partial_{\varphi} & 0 & 0 
\end{array}\right)Y_{\ell m},\\
\mathbf{T}^{(Bt)}_{\ell m}
&=\left(\begin{array}{cccc}
0 & 0 & \csc\theta\partial_{\varphi} & -\sin\theta\partial_{\theta}\\
0 & 0 & 0 & 0\\
\csc\theta\partial_{\varphi} & 0 & 0 & 0 \\
-\sin\theta\partial_{\theta} & 0 & 0 & 0 
\end{array}\right)Y_{\ell m},\\
\mathbf{T}^{(B1)}_{\ell m}
&=\left(\begin{array}{cccc}
0 & 0 & 0 & 0\\
0 & 0 & \csc\theta\partial_{\varphi} & -\sin\theta\partial_{\theta}\\
0 & \csc\theta\partial_{\varphi} & 0 & 0 \\
1 & -\sin\theta\partial_{\theta} & 0 & 0 
\end{array}\right)Y_{\ell m},\\
\mathbf{T}^{(E2)}_{\ell m}
&=\left(\begin{array}{cccc}
0 & 0 & 0 & 0\\
0 & 0 & 0 & 0\\
0 & 0 & \hat{\bm{\mathcal{W}}} & \hat{\bm{\mathcal{V}}}\\
0 & 0 & \hat{\bm{\mathcal{V}}} & -\sin^2\theta\hat{\bm{\mathcal{W}}}
\end{array}\right)Y_{\ell m},\\
\mathbf{T}^{(B2)}_{\ell m}
&=\left(\begin{array}{cccc}
0 & 0 & 0 & 0\\
0 & 0 & 0 & 0\\
0 & 0 & -\csc\theta\hat{\bm{\mathcal{V}}} & \sin\theta\hat{\bm{\mathcal{W}}} \\
0 & 0 & \sin\theta\hat{\bm{\mathcal{W}}} & \sin\theta\hat{\bm{\mathcal{V}}}
\end{array}\right)Y_{\ell m},
\end{align}
\end{subequations}
where
\begin{subequations}
\begin{align}
\hat{\bm{\mathcal{V}}}&=2\partial_{\theta}\partial_{\varphi}-2\cot\theta\partial_{\varphi},\\
\hat{\bm{\mathcal{W}}}&=\partial_{\theta}^2-\cot\theta\partial_{\theta}-\csc^2\theta\partial_{\varphi}^2.
\end{align}
\end{subequations}
These ten basis tensors are divided into two groups: the parity-odd sector, $a\in\{Bt,B1,B2\}$, and the parity-even sector, $a\in\{tt,Rt,L0,T0,Et,E1,E2\}$.

This tensor basis satisfies the orthogonality relation
\begin{equation}
\int\eta^{\mu\alpha}\eta^{\nu\beta}\left[\mathbf{T}^{(a)}_{\ell m}\right]_{\mu\nu}\left\{\left[\mathbf{T}^{(a')}_{\ell'm'}\right]_{\alpha\beta}\right\}^*\dd{\Omega}=\epsilon_{(a)}\delta_{(a)(a')}\delta_{\ell\ell'}\delta_{mm'},    
\end{equation}
where the normalization constants are given by
\begin{equation}
\label{eq_A:normalization_constants}
\begin{aligned}
&\epsilon_{(tt)}=\epsilon_{(L0)}=1,\quad
\epsilon_{(Rt)}=-2,\quad
\epsilon_{(T0)}=2/r^4,\\
&\epsilon_{(Et)}=-\epsilon_{(E1)}=\epsilon_{(Bt)}=-\epsilon_{(B1)}
=-2\ell(\ell+1)/r^2,\\
&\epsilon_{(E2)}=\epsilon_{(B2)}
=2\sigma_{\ell}/r^4.
\end{aligned}
\end{equation}

\section{\label{app:gauge_transformation}Gauge transformation}
To investigate the gauge transformation of GW metric, the gauge vector $\xi_{\alpha}(t,\bm{r})$ is decomposed in terms of vector harmonics, defined as \cite{Maggiore_2018_b}
\begin{subequations}
\begin{align}
\mathbf{V}^{(t)}_{\ell m}
&=(1,0,0,0)Y_{\ell m},\\
\mathbf{V}^{(R)}_{\ell m}
&=(0,\hat{\bm{n}})Y_{\ell m},\\
\mathbf{V}^{(E)}_{\ell m}
&=(0,\bm{\nabla})Y_{\ell m},\\
\mathbf{V}^{(B)}_{\ell m}
&=(i/r)(0,\hat{\bm{L}})Y_{\ell m},
\end{align}
\end{subequations}
where $\hat{n}_{i}=(\sin\theta\cos\varphi,\sin\theta\sin\varphi,\cos\theta)$ is the unit directional vector, $\bm{\nabla}=(\partial_{x},\partial_{y},\partial_{z})$ is the gradient operator, and $\hat{\bm{L}}\equiv-ir\hat{\bm{n}}\times\bm{\nabla}$ is the orbital angular momentum operator. The gauge vector can then be expanded as \cite{Maggiore_2018_b}
\begin{equation}
\xi_{\alpha}(t,\bm{r})
=\frac{1}{2\pi}\int_{-\infty}^{\infty}e^{-ikt}\dd k
\sum_{a\ell m}
\tilde{\xi}^{(a)}_{\ell m}(k,r)
\left[\mathbf{V}_{\ell m}^{(a)}(\theta,\varphi)\right]_{\alpha},
\end{equation}
where the angular quantum number ranges in $\ell\geqslant0$ for $(t)$ and $(R)$ modes, and in $\ell\geqslant1$ for $(E)$ and $(B)$ modes.
The gauge transformation (\ref{eq_2:gauge_transformation}) can be rewritten in terms of the harmonic decomposition as \cite{Maggiore_2018_b}
\begin{subequations}
\label{eq_B:gauge_transformation}
\begin{align}
\tilde{h}_{\ell m}^{(tt)}&\rightarrow\tilde{h}_{\ell m}^{(tt)}
+2ik\tilde{\xi}_{\ell m}^{(t)}
+f'(r)f(r)\tilde{\xi}_{\ell m}^{(R)},\\
\tilde{h}_{\ell m}^{(Rt)}&\rightarrow \tilde{h}_{\ell m}^{(Rt)}
+ik\tilde{\xi}_{\ell m}^{(R)}
-\partial_{r}\tilde{\xi}_{\ell m}^{(t)}+f'(r)f(r)\tilde{\xi}_{\ell m}^{(t)},\\
\tilde{h}_{\ell m}^{(L0)}&\rightarrow \tilde{h}_{\ell m}^{(L0)}
+2ik\tilde{\xi}_{\ell m}^{(R)}
-f'(r)f(r)\tilde{\xi}_{\ell m}^{(R)}\\
\tilde{h}_{\ell m}^{(T0)}&\rightarrow \tilde{h}_{\ell m}^{(T0)}
-2rf(r)\tilde{\xi}_{\ell m}^{(R)}+\ell(\ell+1)\tilde{\xi}_{\ell m}^{(E)},\\
\tilde{h}_{\ell m}^{(Et)}&\rightarrow \tilde{h}_{\ell m}^{(Et)}
-\tilde{\xi}_{\ell m}^{(t)}+ik\tilde{\xi}_{\ell m}^{(E)},\\
\tilde{h}_{\ell m}^{(E1)}&\rightarrow \tilde{h}_{\ell m}^{(E1)}
-(\partial_{r}-2/r)\tilde{\xi}_{\ell m}^{(E)}
-\tilde{\xi}_{\ell m}^{(R)},\\
\tilde{h}_{\ell m}^{(Bt)}&\rightarrow \tilde{h}_{\ell m}^{(Bt)}
-ik\tilde{\xi}_{\ell m}^{(B)}\\
\tilde{h}_{\ell m}^{(B1)}&\rightarrow \tilde{h}_{\ell m}^{(B1)}
+(\partial_{r}-2/r)\tilde{\xi}_{\ell m}^{(B)},\\
\tilde{h}_{\ell m}^{(E2)}&\rightarrow \tilde{h}_{\ell m}^{(E2)}
-\tilde{\xi}_{\ell m}^{(E)},\\
\tilde{h}_{\ell m}^{(B2)}&\rightarrow \tilde{h}_{\ell m}^{(B2)}
-\tilde{\xi}_{\ell m}^{(B)}.
\end{align}
\end{subequations}
The Regge–Wheeler (RW) gauge is achieved through the following procedure \cite{Maggiore_2018_b}. (1) One first appropriately chooses $\tilde{\xi}^{(B)}_{\ell m}$ for $\ell\geqslant2$ such that $\tilde{h}^{(B2)}_{\ell m}=0$ for all $\ell$. (2) One then further fixes $\tilde{\xi}^{(B)}_{1m}$ so that $\tilde{h}^{(B1)}_{1m}=0$. As a result, among odd-parity perturbations, only
$\tilde{h}^{(Bt)}_{\ell m}$ and $\tilde{h}^{(B1)}_{\ell m}$ (with $\ell\geqslant1$) remain non-vanishing. The treatment of even-parity perturbations proceeds analogously. (1) First, one chooses $\tilde{\xi}^{(E)}_{\ell m}$ for $\ell\geqslant2$ such that $\tilde{h}^{(E2)}_{\ell m}=0$ for all $\ell$. (2) Next, one fixes $\tilde{\xi}^{(R)}_{\ell m}$ for $\ell\geqslant1$ so that $\tilde{h}^{(E1)}_{\ell m}=0$. (3) One then chooses $\tilde{\xi}^{(t)}_{\ell m} (\ell\geqslant1)$ such that $\tilde{h}^{(Et)}_{\ell}=0$. 
At this stage, $\tilde{\xi}^{(E)}_{1m}$, $\tilde{\xi}^{(R)}_{00}$, and $\tilde{\xi}^{(t)}_{00}$ remain arbitrary, allowing the elimination of three additional degrees of freedom. (4) Consider first $\tilde{h}^{(T0)}_{\ell m}$. Since $\tilde{\xi}_{00}^{(E)}$ does not exist, one can choose $\tilde{\xi}^{(R)}_{00}$ such that $\tilde{h}^{(T0)}_{00}=0$. (5) Because $\tilde{\xi}^{(R)}_{1m}$ has already been fixed through $\tilde{\xi}^{(E)}_{1m}$, an appropriate choice of $\tilde{\xi}^{(E)}_{1m}$ ensures that $\tilde{h}^{(T0)}_{1m}=0$. (6) Finally, the remaining gauge function $\tilde{\xi}^{(t)}_{00}$ can be chosen appropriately to impose $\tilde{h}^{(Rt)}_{00}=0$.

\section{\label{app:wave_field}Wave fields of apparent polarizations}
The wave fields of the apparent polarizations of scattered GWs, as defined in Eq.\,(\ref{eq_4:GW_polarizations}), are displayed in Fig.\,\ref{fig:wave_field_apparent_modes}. Compared with the $+$ and $\times$ modes, these apparent modes generally have smaller amplitudes, and no Poisson spot is formed. This behavior arises from the axisymmetric nature of the scattering process considered here.
\begin{figure*}[htbp]
    \centering
    \includegraphics[width=1.0\linewidth]{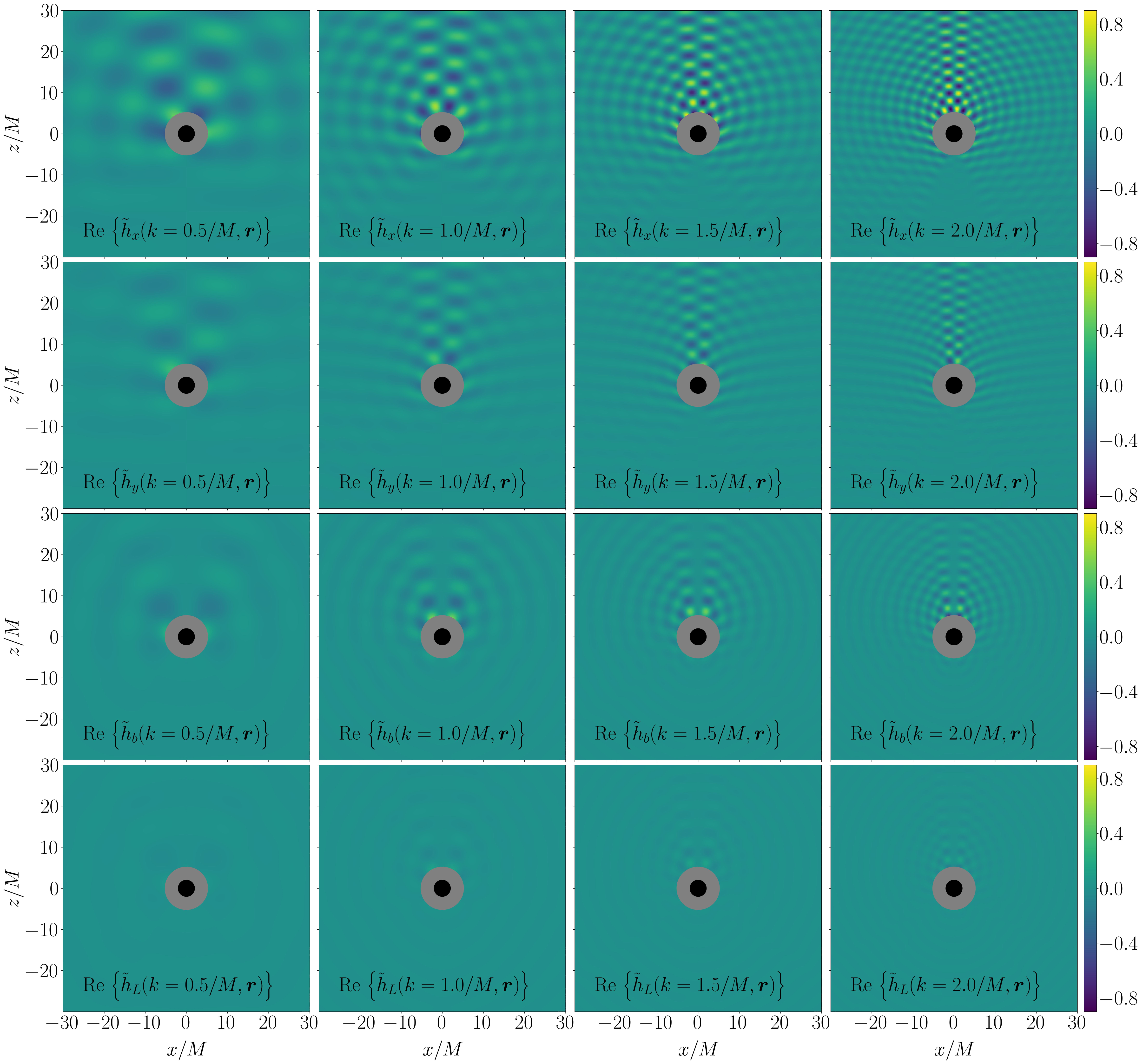}
    \caption{Wave fields of the apparent polarizations, namely the $x$, $y$, $b$, and $L$ modes, for the scattered GWs. The black and gray regions represent the BH event horizon and the light ring, respectively.}
    \label{fig:wave_field_apparent_modes}
\end{figure*}

\section{\label{app:conventional_computations}Overview on the conventional computations with asymptotic expansion}
In the BH perturbation theory, the GW polarization modes can be written as \cite{Martel_2005,Bao_2022}
\begin{subequations}
\label{eq_app:GW_polarization}
\begin{align}
\tilde{h}_{+}
&=\frac{1}{r}\sum_{\ell=2}^{\infty}
\sum_{m=-\ell}^{\ell}
\left\{\tilde{\psi}^{(+)}_{\ell m}(r)
\left[(\partial_{\theta}^2Y_{\ell m})
+\frac{\ell(\ell+1)}{2}Y_{\ell m}\right]-\frac{2m}{k\sin\theta}
\tilde{\psi}_{\ell m}^{(-)}(r)
\left[(\partial_{\theta}Y_{\ell m})
-\cot\theta Y_{\ell m}\right]\right\},\\
i\tilde{h}_{\times}
&=\frac{1}{r}\sum_{\ell=2}^{\infty}
\sum_{m=-\ell}^{\ell}
\left\{\frac{2}{k}
\tilde{\psi}_{\ell m}^{(-)}(r)
\left[(\partial_{\theta}^2Y_{\ell m})
+\frac{\ell(\ell+1)}{2}Y_{\ell m}\right]-\frac{m}{\sin\theta}
\tilde{\psi}^{(+)}_{\ell m}(r)
\left[(\partial_{\theta}Y_{\ell m})
-\cot\theta Y_{\ell m}\right]\right\},
\end{align}
\end{subequations}
for the observer who is located sufficiently distant from the BH. Here, $\tilde{\psi}_{\ell m}^{(\pm)}$ are the parity-even/odd master functions. Substituting their asymptotic expansion (\ref{eq_2:boundary_conditions_inf}) into Eq.\,(\ref{eq_app:GW_polarization}), one finds that the full GW consists of a distorted plane wave and a reflected spherical wave, namely,
\begin{equation}
\tilde{h}_{+,\times}=\Big\{\text{distorted plane wave}\Big\}_{+,\times}
+\tilde{h}_{+,\times}^{\rm(Ref)}.
\end{equation}
The plane-wave term itself is also expressed as a divergent partial-wave series. In previous studies, e.g., \cite{Pijnenburg_2024b,Chan_2025}, it is usually approximated by the incident plane GW, $\tilde{h}^{(0)}_{+,\times}$, given in Eq.\,(\ref{eq_5:incident_polarization}).

In terms of the left- and right-handed polarizations, the reflected spherical wave is written as
\begin{equation}
\left(\begin{array}{@{}c@{}}
\tilde{h}_{L}^{\rm(Ref)}\\[6pt]
\tilde{h}_{R}^{\rm(Ref)}
\end{array}\right)
=\frac{e^{ikr_{*}}}{r}\times\bm{\mathcal{M}}
\left(\begin{array}{@{}c@{}}
\tilde{\mathcal{A}}_L\\[6pt]
\tilde{\mathcal{A}}_R
\end{array}\right),
\end{equation}
where $e^{ikr{}}/r$ represents the radial dependence of spherical waves, and $\bm{\mathcal{M}}$ encodes the polarization conversion during scattering at different angles. The diagonal elements of $\bm{\mathcal{M}}$ describe parity-even scattering processes, while the off-diagonal elements correspond to parity-odd scattering, leading to conservation and mixing of the initial polarization modes. The explicit form of this matrix is
\begin{equation}
\label{eq_app:conservation_matrix}
\bm{\mathcal{M}}
\equiv\frac{\pi}{ik}
\left(\begin{array}{@{}cc@{}}
\mathcal{L}_{1}\mathcal{L}_0\Big\{f_{s}(\theta)e^{-2i\varphi}\Big\} &
\mathcal{L}_{1}\mathcal{L}_0\Big\{f_{a}(\theta)e^{2i\varphi}\Big\}\\[9pt]
-\bar{\mathcal{L}}_{-1}\bar{\mathcal{L}}_0\Big\{f_{a}(\theta)e^{-2i\varphi}\Big\}&
\bar{\mathcal{L}}_{-1}\bar{\mathcal{L}}_0\Big\{f_{s}(\theta)e^{2i\varphi}\Big\}
\end{array}\right),
\end{equation}
The symmetric and antisymmetric angular factors, $f_{s}(\theta)$ and $f_{a}(\theta)$, are defined as
\begin{subequations}
\label{eq_app:partial_wave_series}
\begin{align}
f_{s}(\theta)&\equiv\sum_{\ell=2}^{\infty}\sum_{\mathrm{p}=\pm1}\left(\frac{2\ell+1}{4\pi\sigma_{\ell}}\right)\left[e^{2i\delta_{\ell}^{\rm(p)}}-1\right]\mathrm{P}_{\ell2}(\cos\theta),\\
f_{a}(\theta)&\equiv\sum_{\ell=2}^{\infty}\sum_{\mathrm{p}=\pm1}\mathrm{p}\left(\frac{2\ell+1}{4\pi\sigma_{\ell}}\right)\left[e^{2i\delta_{\ell}^{\rm(p)}}-1\right]\mathrm{P}_{\ell2}(\cos\theta),
\end{align}
\end{subequations}
where $\mathrm{P}_{\ell m}(\cos\theta)$ denotes the associated Legendre function of the first kind. As discussed in the main text, the series appearing in the conservation matrix (\ref{eq_app:conservation_matrix}) are divergent. The standard treatment is therefore to introduce a regularization procedure and replace $\bm{\mathcal{M}}$ with a regularized matrix $\widehat{\bm{\mathcal{M}}}$, namely,
\begin{equation}
\widehat{\bm{\mathcal{M}}}
=\frac{\pi}{ik}
\left(\begin{array}{@{}cc@{}}
\mathcal{L}_{1}\mathcal{L}_0\Big\{\widehat{f}_{s}(\theta)e^{-2i\varphi}\Big\} &
\mathcal{L}_{1}\mathcal{L}_0\Big\{\widehat{f}_{a}(\theta)e^{2i\varphi}\Big\}\\[9pt]
-\bar{\mathcal{L}}_{-1}\bar{\mathcal{L}}_0\Big\{\widehat{f}_{a}(\theta)e^{-2i\varphi}\Big\}&
\bar{\mathcal{L}}_{-1}\bar{\mathcal{L}}_0\Big\{\widehat{f}_{s}(\theta)e^{2i\varphi}\Big\}
\end{array}\right),
\end{equation}
where $\widehat{f}_{s}(\theta)$ and $\widehat{f}_{a}(\theta)$ are the regularized series corresponding to Eq.\,(\ref{eq_app:partial_wave_series}). As a representative example, the series reduction method is briefly reviewed in Appendix \ref{app:series_reduction_method} and applied in our calculation to regularize the divergences in $\widehat{f}_{s}(\theta)$ and $\widehat{f}_{a}(\theta)$.

\begin{figure*}[htbp]
    \centering
    \includegraphics[width=1.0\linewidth]{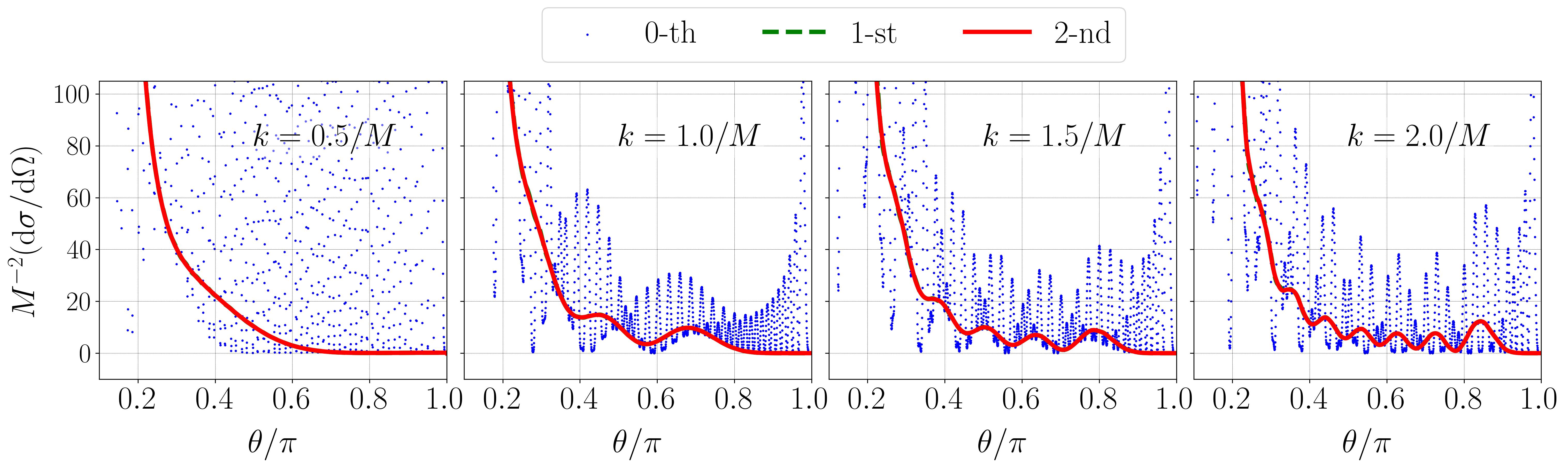}
    \caption{Differential cross sections calculated from Eq.\,(\ref{eq_app:differential_cross_section}) for $k=\{0.5,1.0,1.5,2.0\}/M$. The blue dots show the resummation without regularization, while the green and red curves correspond to the results obtained using first- and second-order series reduction, respectively.}
    \label{fig:differential_cross_section}
\end{figure*}

\section{\label{app:series_reduction_method}Series reduction method}
The series reduction method was developed to address the divergence of the series involving the Legendre function \cite{Yennie_1954,Stratton_2020}, which are generally written as
\begin{equation}
\label{eq_app:f}
f(\theta)=\sum_{\ell=2}^{\infty}
a_{\ell}\mathrm{P}_{\ell2}(\cos\theta),
\end{equation}
where $a_{\ell}$ denotes the original coefficient. The $k$-th reduced series is defined by
\begin{equation}
\label{eq_app:reduced_f}
f^{(k)}(\theta)
=(1-\cos\theta)^{k}f(\theta)
=\sum_{\ell=2}^{\infty}a_{\ell}(1-\cos\theta)^{k}{\rm P}_{\ell2}(\cos\theta).
\end{equation}
Applying the identity
\begin{equation}
(1-\cos\theta)\mathrm{P}_{\ell2}(\cos\theta)=\mathrm{P}_{\ell2}(\cos\theta)-\frac{\ell-1}{2\ell+1}\mathrm{P}_{\ell+1,2}(\cos\theta)-\frac{\ell+2}{2\ell+1}\mathrm{P}_{\ell-1,2}(\cos\theta),
\end{equation}
to the right-hand side of Eq.\,(\ref{eq_app:reduced_f}), one finds that the reduced series can be rewritten as
\begin{equation}
f^{(k)}(\theta)
=\sum_{\ell=2}^{\infty}a^{(k)}_{\ell}{\rm P}_{\ell2}(\cos\theta),
\end{equation}
which has the same form as Eq.\,(\ref{eq_app:f}). Here,  $a^{(k)}_{\ell}$ is the $k$-th reduction coefficient, which satisfy the recursive relation,
\begin{equation}
\label{eq_app:recursive}
a^{(k+1)}_{\ell}
=a^{(k)}_{\ell}
-\frac{\ell-2}{2\ell-1}a^{(k)}_{\ell-1}
-\frac{\ell+3}{2\ell+3}a^{(k)}_{\ell+1}.
\end{equation}
In deriving Eq.\,(\ref{eq_app:recursive}), we have set $a^{(k)}_{1}=0$. This prescription allows the coefficients $a^{(k)}_{\ell}$ to be computed iteratively from the original series. \ then gives the final reduced series 
\begin{equation}
\widehat{f}(\theta)=(1-\cos\theta)^{-k}f^{(k)}(\theta).
\end{equation}
The convergence of the series is significantly improved by this procedure, except in the vicinity of the optical axis.

As an important application, we compute the differential cross section for GWs scattered by a Schwarzschild BH using the regularization scheme described above. Following Ref.\,\cite{Dolan_2008a}, we consider incident GWs with circular polarization, for which $\tilde{\mathcal{A}}_{\times}=i\tilde{\mathcal{A}}_{+}\equiv i\tilde{\mathcal{A}}$. Consequently, $\tilde{\mathcal{A}}_{L}=0$ and $\tilde{\mathcal{A}}_{R}=\sqrt{2}\tilde{\mathcal{A}}$. 
The differential cross section is defined as the ratio of the energy flux carried by the reflected spherical GWs to that of the incident plane GWs,
\begin{equation}
\label{eq_app:differential_cross_section}
\frac{\dd\sigma}{\dd\Omega}
=r^2\frac{\frac{k^2}{32\pi}\left\{ |h_{L}^{\rm(Ref)}|^2+|h_{R}^{\rm(Ref)}|^2\right\}}{\frac{k^2}{32\pi}\left\{|\tilde{\mathcal{A}}_{L}|^2+|\tilde{\mathcal{A}}_{R}|^2\right\}}
=\left|\mathcal{M}_{22}\right|^2+\left|\mathcal{M}_{12}\right|^2.
\end{equation}
Here, the conversion matrix $\bm{\mathcal{M}}$ is defined in Eq.\,(\ref{eq_app:conservation_matrix}), with the element $\mathcal{M}_{22}$ representing the helicity-preserving scattering, while $\mathcal{M}_{12}$ corresponding to the helicity-reversving scattering. Fig.\,\ref{fig:differential_cross_section} displays the numerical results for $k=\{0.5,1.0,1.5,2.0\}/M$, regularized by the series reduction, and exhibits the characteristic singular behavior along the optical axis.

\end{widetext}
\bibliography{reference}

@book{Weinberg,
  title         = {{Gravitation and Cosmology}},
  author        = "S. Weinberg",
  publisher     = "John Wiley \& Sons", 
  address       = "Inc", 
  year          = "1972",
}

@book{MTW,
  title={Gravitation},
  author={Misner, Charles W and Thorne, Kip S and Wheeler, John Archibald},
  year={1973},
  publisher={Macmillan}
}

@article{Wang_2024,
doi = {10.3847/1538-4357/ad6c4d},
url = {https://dx.doi.org/10.3847/1538-4357/ad6c4d},
year = {2024},
month = {oct},
publisher = {The American Astronomical Society},
volume = {974},
number = {1},
pages = {7},
author = {Wang, Qingqing and others},
title = {{Observational Test of $f(Q)$ Gravity with Weak Gravitational Lensing}},
journal = {The Astrophysical Journal},
}

@article{Cao_2012,
doi = {10.1088/0004-637X/755/1/31},
url = {https://dx.doi.org/10.1088/0004-637X/755/1/31},
year = {2012},
month = {jul},
publisher = {The American Astronomical Society},
volume = {755},
number = {1},
pages = {31},
author = {Cao, Shuo and Covone, Giovanni and Zhu, ZongHong},
title = {TESTING THE DARK ENERGY WITH GRAVITATIONAL LENSING STATISTICS},
journal = {The Astrophysical Journal},
}

@article{Yang_2022,
    author = {Yang, Lilan and others},
    title = {The size–luminosity relation of lensed galaxies at z $\sim$ 6–9 in the {Hubble} Frontier Fields},
    journal = {Monthly Notices of the Royal Astronomical Society},
    volume = {514},
    number = {1},
    pages = {1148-1161},
    year = {2022},
    month = {06},
    issn = {0035-8711},
    doi = {10.1093/mnras/stac1236},
    url = {https://doi.org/10.1093/mnras/stac1236},
}

@article{Dan_2012,
  title={{CLASH}: three strongly lensed images of a candidate z$\approx$11 galaxy},
  author={Coe, Dan and others},
  journal={The Astrophysical Journal},
  volume={762},
  number={1},
  pages={32},
  year={2012},
  publisher={IOP Publishing}
}

@article{Bolton_2008,
doi = {10.1086/589327},
url = {https://doi.org/10.1086/589327},
year = {2008},
month = {aug},
publisher = {},
volume = {682},
number = {2},
pages = {964},
author = {Bolton, Adam S. and others},
title = {{The Sloan Lens ACS Survey. V. The Full ACS Strong-Lens Sample}},
journal = {The Astrophysical Journal}
}

@article{Refregier_2003,
   author = {Refregier, Alexandre},
   title = {Weak Gravitational Lensing by Large-Scale Structure}, 
   journal= {Annual Review of Astronomy and Astrophysics},
   year = {2003},
   volume = {41},
   number = {Volume 41, 2003},
   pages = {645-668},
   doi = {https://doi.org/10.1146/annurev.astro.41.111302.102207},
   publisher = {Annual Reviews},
   issn = {1545-4282},
   type = {Journal Article},
  }

@article{Liu_2022,
doi = {10.3847/1538-4357/ac4c3b},
url = {https://dx.doi.org/10.3847/1538-4357/ac4c3b},
year = {2022},
month = {mar},
publisher = {The American Astronomical Society},
volume = {927},
number = {1},
pages = {28},
author = {Liu, XiaoHui and others},
title = {Galaxy-scale Test of General Relativity with Strong Gravitational Lensing},
journal = {The Astrophysical Journal},
}

@article{Piorkowska_2013,
doi = {10.1088/1475-7516/2013/10/022},
url = {https://dx.doi.org/10.1088/1475-7516/2013/10/022},
year = {2013},
month = {oct},
publisher = {},
volume = {2013},
number = {10},
pages = {022},
author = {Aleksandra Piórkowska and Marek Biesiada and Zonghong Zhu},
title = {{Strong gravitational lensing of gravitational waves in Einstein Telescope}},
journal = {Journal of Cosmology and Astroparticle Physics},
}

@article{Lin_2023,
  title = {{Detecting strong gravitational lensing of gravitational waves with TianQin}},
  author = {Lin, XinYi and others},
  journal = {Phys. Rev. D},
  volume = {108},
  issue = {6},
  pages = {064020},
  numpages = {12},
  year = {2023},
  month = {Sep},
  publisher = {American Physical Society},
  doi = {10.1103/PhysRevD.108.064020},
  url = {https://link.aps.org/doi/10.1103/PhysRevD.108.064020}
}

@article{Ezquiaga_2021,
  title = {Phase effects from strong gravitational lensing of gravitational waves},
  author = {Ezquiaga, Jose Mar\'{\i}a and others},
  journal = {Phys. Rev. D},
  volume = {103},
  issue = {6},
  pages = {064047},
  numpages = {28},
  year = {2021},
  month = {Mar},
  publisher = {American Physical Society},
  doi = {10.1103/PhysRevD.103.064047},
  url = {https://link.aps.org/doi/10.1103/PhysRevD.103.064047}
}

@article{Takahashi_2003,
doi = {10.1086/377430},
url = {https://dx.doi.org/10.1086/377430},
year = {2003},
month = {oct},
publisher = {},
volume = {595},
number = {2},
pages = {1039},
author = {Takahashi, Ryuichi and Nakamura, Takashi},
title = {Wave Effects in the Gravitational Lensing of Gravitational Waves from Chirping Binaries},
journal = {The Astrophysical Journal},
}

@article{Meena_2019,
    author = {Meena, Ashish Kumar and Bagla, Jasjeet Singh},
    title = {Gravitational lensing of gravitational waves: wave nature and prospects for detection},
    journal = {Monthly Notices of the Royal Astronomical Society},
    volume = {492},
    number = {1},
    pages = {1127-1134},
    year = {2019},
    month = {12},
    issn = {0035-8711},
    doi = {10.1093/mnras/stz3509},
    url = {https://doi.org/10.1093/mnras/stz3509},
}

@article{Nakamura_1998,
  title = {Gravitational Lensing of Gravitational Waves from Inspiraling Binaries by a Point Mass Lens},
  author = {Nakamura, Takahiro T.},
  journal = {Phys. Rev. Lett.},
  volume = {80},
  issue = {6},
  pages = {1138--1141},
  numpages = {0},
  year = {1998},
  month = {Feb},
  publisher = {American Physical Society},
  doi = {10.1103/PhysRevLett.80.1138},
  url = {https://link.aps.org/doi/10.1103/PhysRevLett.80.1138}
}

@article{Haris_2018,
  title={Identifying strongly lensed gravitational wave signals from binary black hole mergers},
  author={Haris, K and others},
  journal={arXiv preprint arXiv:1807.07062},
  year={2018}
}

@article{Hannuksela_2019,
doi = {10.3847/2041-8213/ab0c0f},
url = {https://dx.doi.org/10.3847/2041-8213/ab0c0f},
year = {2019},
month = {mar},
publisher = {The American Astronomical Society},
volume = {874},
number = {1},
pages = {L2},
author = {O. A. Hannuksela and others},
title = {{Search for Gravitational Lensing Signatures in LIGO-Virgo Binary Black Hole Events}},
journal = {The Astrophysical Journal Letters},
}

@article{McIsaac_2020,
  title = {{Search for strongly lensed counterpart images of binary black hole mergers in the first two LIGO observing runs}},
  author = {McIsaac, Connor and others},
  journal = {Phys. Rev. D},
  volume = {102},
  issue = {8},
  pages = {084031},
  numpages = {16},
  year = {2020},
  month = {Oct},
  publisher = {American Physical Society},
  doi = {10.1103/PhysRevD.102.084031},
  url = {https://link.aps.org/doi/10.1103/PhysRevD.102.084031}
}

@article{XiaoshuLiu_2021,
doi = {10.3847/1538-4357/abd7eb},
url = {https://dx.doi.org/10.3847/1538-4357/abd7eb},
year = {2021},
month = {feb},
publisher = {The American Astronomical Society},
volume = {908},
number = {1},
pages = {97},
author = {Xiaoshu Liu and Ignacio Magaña Hernandez and Jolien Creighton},
title = {{Identifying Strong Gravitational-wave Lensing during the Second Observing Run of Advanced LIGO and Advanced Virgo}},
journal = {The Astrophysical Journal},
}

@article{LIGO_2021,
doi = {10.3847/1538-4357/ac23db},
url = {https://dx.doi.org/10.3847/1538-4357/ac23db},
year = {2021},
month = {dec},
publisher = {The American Astronomical Society},
volume = {923},
number = {1},
pages = {14},
author = {R. Abbott and others},
collaboration = {The LIGO Scientific Collaboration and the Virgo Collaboration},
title = {{Search for Lensing Signatures in the Gravitational-Wave Observations from the First Half of LIGO–Virgo’s Third Observing Run}},
journal = {The Astrophysical Journal},
}

@article{Lo_2023,
  title = {Bayesian statistical framework for identifying strongly lensed gravitational-wave signals},
  author = {Lo, Rico K. L. and Maga\~na Hernandez, Ignacio},
  journal = {Phys. Rev. D},
  volume = {107},
  issue = {12},
  pages = {123015},
  numpages = {26},
  year = {2023},
  month = {Jun},
  publisher = {American Physical Society},
  doi = {10.1103/PhysRevD.107.123015},
  url = {https://link.aps.org/doi/10.1103/PhysRevD.107.123015}
}

@article{Abbott_2024,
doi = {10.3847/1538-4357/ad3e83},
url = {https://doi.org/10.3847/1538-4357/ad3e83},
year = {2024},
month = {jul},
publisher = {The American Astronomical Society},
volume = {970},
number = {2},
pages = {191},
author = {Abbott, R. and others},
title = {{Search for Gravitational-lensing Signatures in the Full Third Observing Run of the LIGO–Virgo Network}},
journal = {The Astrophysical Journal},
}

@article{Vieira_2013,
  title={Dusty starburst galaxies in the early Universe as revealed by gravitational lensing},
  author={Vieira, JD and others},
  journal={Nature},
  volume={495},
  number={7441},
  pages={344--347},
  year={2013},
  publisher={Nature Publishing Group UK London}
}

@article{Lo_2025,
  title = {Observational Signatures of Highly Magnified Gravitational Waves from Compact Binary Coalescence},
  author = {Lo, Rico K. L. and others},
  journal = {Phys. Rev. Lett.},
  volume = {134},
  issue = {15},
  pages = {151401},
  numpages = {7},
  year = {2025},
  month = {Apr},
  publisher = {American Physical Society},
  doi = {10.1103/PhysRevLett.134.151401},
  url = {https://link.aps.org/doi/10.1103/PhysRevLett.134.151401}
}

@book{Maggiore_2018_b,
    author = {Maggiore, Michele},
    title = {Gravitational Waves: Volume 2: Astrophysics and Cosmology},
    publisher = {Oxford University Press},
    year = {2018},
    month = {03},
    isbn = {9780198570899},
    doi = {10.1093/oso/9780198570899.001.0001},
    url = {https://doi.org/10.1093/oso/9780198570899.001.0001},
}

@book{Maggiore_2018_a,
    author = {Maggiore, Michele},
    title = {Gravitational Waves: Volume 1: Theory and Experiments},
    publisher = {Oxford University Press},
    year = {2007},
    month = {10},
    isbn = {9780198570745},
    doi = {10.1093/acprof:oso/9780198570745.001.0001},
    url = {https://doi.org/10.1093/acprof:oso/9780198570745.001.0001},
}

@article{Sereno_2006,
  title = {Analytical {Kerr} black hole lensing in the weak deflection limit},
  author = {Sereno, Mauro and De Luca, Fabiana},
  journal = {Phys. Rev. D},
  volume = {74},
  issue = {12},
  pages = {123009},
  numpages = {11},
  year = {2006},
  month = {Dec},
  publisher = {American Physical Society},
  doi = {10.1103/PhysRevD.74.123009},
  url = {https://link.aps.org/doi/10.1103/PhysRevD.74.123009}
}

@article{Keeton_2005,
  title = {Formalism for testing theories of gravity using lensing by compact objects: Static, spherically symmetric case},
  author = {Keeton, Charles R. and Petters, A. O.},
  journal = {Phys. Rev. D},
  volume = {72},
  issue = {10},
  pages = {104006},
  numpages = {16},
  year = {2005},
  month = {Nov},
  publisher = {American Physical Society},
  doi = {10.1103/PhysRevD.72.104006},
  url = {https://link.aps.org/doi/10.1103/PhysRevD.72.104006}
}

@article{Guo_2020,
  title = {Convergence and efficiency of different methods to compute the diffraction integral for gravitational lensing of gravitational waves},
  author = {Guo, Xiao and Lu, Youjun},
  journal = {Phys. Rev. D},
  volume = {102},
  issue = {12},
  pages = {124076},
  numpages = {15},
  year = {2020},
  month = {Dec},
  publisher = {American Physical Society},
  doi = {10.1103/PhysRevD.102.124076},
  url = {https://link.aps.org/doi/10.1103/PhysRevD.102.124076}
}

@article{Baraldo_1999,
  title = {Gravitationally induced interference of gravitational waves by a rotating massive object},
  author = {Baraldo, Christian and Hosoya, Akio and Nakamura, Takahiro T.},
  journal = {Phys. Rev. D},
  volume = {59},
  issue = {8},
  pages = {083001},
  numpages = {8},
  year = {1999},
  month = {Mar},
  publisher = {American Physical Society},
  url = {https://link.aps.org/doi/10.1103/PhysRevD.59.083001}
}

@article{Pijnenburg_2024a,
    author = {Pijnenburg, Martin and others},
    title = {Rutherford scattering of quantum and classical fields},
    journal = {American Journal of Physics},
    volume = {92},
    number = {8},
    pages = {597-605},
    year = {2024},
    month = {08},
    issn = {0002-9505},
    doi = {10.1119/5.0175025},
    url = {https://doi.org/10.1119/5.0175025},
}

@article{Regge_Wheeler_1957,
  title = {Stability of a {Schwarzschild} Singularity},
  author = {Regge, Tullio and Wheeler, John A.},
  journal = {Phys. Rev.},
  volume = {108},
  issue = {4},
  pages = {1063--1069},
  numpages = {0},
  year = {1957},
  month = {Nov},
  publisher = {American Physical Society},
  doi = {10.1103/PhysRev.108.1063},
  url = {https://link.aps.org/doi/10.1103/PhysRev.108.1063}
}

@article{Zerilli_1970b,
  title = {Effective Potential for Even-Parity Regge-Wheeler Gravitational Perturbation Equations},
  author = {Zerilli, Frank J.},
  journal = {Phys. Rev. Lett.},
  volume = {24},
  issue = {13},
  pages = {737--738},
  numpages = {0},
  year = {1970},
  month = {Mar},
  publisher = {American Physical Society},
  doi = {10.1103/PhysRevLett.24.737},
  url = {https://link.aps.org/doi/10.1103/PhysRevLett.24.737}
}

@article{Zerilli_1970a,
  title = {Gravitational Field of a Particle Falling in a {Schwarzschild} Geometry Analyzed in Tensor Harmonics},
  author = {Zerilli, Frank J.},
  journal = {Phys. Rev. D},
  volume = {2},
  issue = {10},
  pages = {2141--2160},
  numpages = {0},
  year = {1970},
  month = {Nov},
  publisher = {American Physical Society},
  doi = {10.1103/PhysRevD.2.2141},
  url = {https://link.aps.org/doi/10.1103/PhysRevD.2.2141}
}

@ARTICLE{Teukolsky_1973,
       author = {{Teukolsky}, Saul A.},
        title = "{Perturbations of a Rotating Black Hole. I. Fundamental Equations for Gravitational, Electromagnetic, and Neutrino-Field Perturbations}",
      journal = {\apj},
         year = 1973,
        month = {oct},
       volume = {185},
        pages = {635-648},
          doi = {10.1086/152444},
       adsurl = {https://ui.adsabs.harvard.edu/abs/1973ApJ...185..635T}
}

@ARTICLE{Teukolsky_1974,
       author = {{Teukolsky}, S.~A. and {Press}, W.~H.},
        title = "{Perturbations of a rotating black hole. III. Interaction of the hole with gravitational and electromagnetic radiation.}",
      journal = {\apj},
         year = {1974},
        month = {oct},
       volume = {193},
        pages = {443-461},
          doi = {10.1086/153180},
       adsurl = {https://ui.adsabs.harvard.edu/abs/1974ApJ...193..443T}
}

@ARTICLE{Press_1973,
       author = {{Press}, William H. and {Teukolsky}, Saul A.},
        title = "{Perturbations of a Rotating Black Hole. II. Dynamical Stability of the Kerr Metric}",
      journal = {\apj},
         year = {1973},
        month = {oct},
       volume = {185},
        pages = {649-674},
          doi = {10.1086/152445},
       adsurl = {https://ui.adsabs.harvard.edu/abs/1973ApJ...185..649P}
}

@article{Bardeen_1973,
    author = {Bardeen, James M. and Press, William H.},
    title = {Radiation fields in the {Schwarzschild} background},
    journal = {Journal of Mathematical Physics},
    volume = {14},
    number = {1},
    pages = {7-19},
    year = {1973},
    month = {01},
    issn = {0022-2488},
    doi = {10.1063/1.1666175},
    url = {https://doi.org/10.1063/1.1666175},
}

@article{Chandrasekhar_1978a,
author = {Chandrasekhar, Subrahmanyan },
title = {The gravitational perturbations of the {Kerr} black hole I. The perturbations in the quantities which vanish in the stationary state},
journal = {Proceedings of the Royal Society of London. A. Mathematical and Physical Sciences},
volume = {358},
number = {1695},
pages = {421-439},
year = {1978},
doi = {10.1098/rspa.1978.0020},
URL = {https://royalsocietypublishing.org/doi/abs/10.1098/rspa.1978.0020},
}

@article{Chandrasekhar_1978b,
author = {Chandrasekhar, Subrahmanyan },
title = {The gravitational perturbations of the {Kerr} black hole. II. The perturbations in the quantities which are finite in the stationary state},
journal = {Proceedings of the Royal Society of London. A. Mathematical and Physical Sciences},
volume = {358},
number = {1695},
pages = {441-465},
year = {1978},
doi = {10.1098/rspa.1978.0021},
URL = {https://royalsocietypublishing.org/doi/abs/10.1098/rspa.1978.0021},
}

@BOOK{Chandrasekhar1983,
       author = {{Chandrasekhar}, S.},
        title = "{The mathematical theory of black holes}",
         year = 1983,
       adsurl = {https://ui.adsabs.harvard.edu/abs/1983mtbh.book.....C}
}

@article{Sasaki_Nakamura_1982,
    author = {Sasaki, Misao and Nakamura, Takashi},
    title = {{Gravitational Radiation from a Kerr Black Hole. I. Formulation and a Method for Numerical Analysis}},
    journal = {Progress of Theoretical Physics},
    volume = {67},
    number = {6},
    pages = {1788-1809},
    year = {1982},
    month = {06},
    issn = {0033-068X},
    doi = {10.1143/PTP.67.1788},
    url = {https://doi.org/10.1143/PTP.67.1788},
}

@article{Dolan_2008a,
doi = {10.1088/0264-9381/25/23/235002},
url = {https://dx.doi.org/10.1088/0264-9381/25/23/235002},
year = {2008},
month = {nov},
publisher = {},
volume = {25},
number = {23},
pages = {235002},
author = {Dolan, Sam R},
title = {Scattering and absorption of gravitational plane waves by rotating black holes},
journal = {Classical and Quantum Gravity},
}

@article{Dolan_2008b,
  title = {Scattering of long-wavelength gravitational waves},
  author = {Dolan, Sam R.},
  journal = {Phys. Rev. D},
  volume = {77},
  issue = {4},
  pages = {044004},
  numpages = {5},
  year = {2008},
  month = {Feb},
  publisher = {American Physical Society},
  doi = {10.1103/PhysRevD.77.044004},
  url = {https://link.aps.org/doi/10.1103/PhysRevD.77.044004}
}

@article{Chrzanowski_1976,
  title = {Zero-mass plane waves in nonzero gravitational backgrounds},
  author = {Chrzanowski, Paul L. and others},
  journal = {Phys. Rev. D},
  volume = {14},
  issue = {2},
  pages = {317--326},
  numpages = {0},
  year = {1976},
  month = {Jul},
  publisher = {American Physical Society},
  doi = {10.1103/PhysRevD.14.317},
  url = {https://link.aps.org/doi/10.1103/PhysRevD.14.317}
}

@article{Kubota_2024_b,
  title = {Spin wave optics for gravitational waves lensed by a {Kerr} black hole},
  author = {Kubota, Keiichiro and others},
  journal = {Phys. Rev. D},
  volume = {110},
  issue = {12},
  pages = {124011},
  numpages = {17},
  year = {2024},
  month = {Dec},
  publisher = {American Physical Society},
  doi = {10.1103/PhysRevD.110.124011},
  url = {https://link.aps.org/doi/10.1103/PhysRevD.110.124011}
}

@article{Glampedakis_2001,
doi = {10.1088/0264-9381/18/10/309},
url = {https://dx.doi.org/10.1088/0264-9381/18/10/309},
year = {2001},
month = {may},
publisher = {},
volume = {18},
number = {10},
pages = {1939},
author = {Kostas Glampedakis and Nils Andersson},
title = {Scattering of scalar waves by rotating black holes},
journal = {Classical and Quantum Gravity},
}

@article{Andersson_1995,
  title = {Scattering of massless scalar waves by a {Schwarzschild} black hole: A phase-integral study},
  author = {Andersson, Nils},
  journal = {Phys. Rev. D},
  volume = {52},
  issue = {4},
  pages = {1808--1820},
  numpages = {0},
  year = {1995},
  month = {Aug},
  publisher = {American Physical Society},
  doi = {10.1103/PhysRevD.52.1808},
  url = {https://link.aps.org/doi/10.1103/PhysRevD.52.1808}
}

@article{Matzner_1968,
    author = {Matzner, Richard A.},
    title = {Scattering of Massless Scalar Waves by a {Schwarzschild} ``Singularity''},
    journal = {Journal of Mathematical Physics},
    volume = {9},
    number = {1},
    pages = {163-170},
    year = {1968},
    month = {01},
    issn = {0022-2488},
    doi = {10.1063/1.1664470},
    url = {https://doi.org/10.1063/1.1664470},
}

@article{Yennie_1954,
  title = {Phase-Shift Calculation of High-Energy Electron Scattering},
  author = {Yennie, D. R. and Ravenhall, D. G. and Wilson, R. N.},
  journal = {Phys. Rev.},
  volume = {95},
  issue = {2},
  pages = {500--512},
  numpages = {0},
  year = {1954},
  month = {Jul},
  publisher = {American Physical Society},
  doi = {10.1103/PhysRev.95.500},
  url = {https://link.aps.org/doi/10.1103/PhysRev.95.500}
}

@article{Stratton_2020,
  title = {Series reduction method for scattering of planar waves by {Kerr} black holes},
  author = {Stratton, Tom and others},
  journal = {Phys. Rev. D},
  volume = {102},
  issue = {4},
  pages = {044025},
  numpages = {10},
  year = {2020},
  month = {Aug},
  publisher = {American Physical Society},
  doi = {10.1103/PhysRevD.102.044025},
  url = {https://link.aps.org/doi/10.1103/PhysRevD.102.044025}
}

@article{Pijnenburg_2024b,
  title = {Wave optics lensing of gravitational waves: Theory and phenomenology of triple systems in the LISA band},
  author = {Pijnenburg, Martin and others},
  journal = {Phys. Rev. D},
  volume = {110},
  issue = {4},
  pages = {044054},
  numpages = {25},
  year = {2024},
  month = {Aug},
  publisher = {American Physical Society},
  doi = {10.1103/PhysRevD.110.044054},
  url = {https://link.aps.org/doi/10.1103/PhysRevD.110.044054}
}

@article{Folacci_2019a,
  title = {Regge pole description of scattering of scalar and electromagnetic waves by a {Schwarzschild} black hole},
  author = {Folacci, Antoine and Hadj, Mohamed Ould El},
  journal = {Phys. Rev. D},
  volume = {99},
  issue = {10},
  pages = {104079},
  numpages = {21},
  year = {2019},
  month = {May},
  publisher = {American Physical Society},
  doi = {10.1103/PhysRevD.99.104079},
  url = {https://link.aps.org/doi/10.1103/PhysRevD.99.104079}
}

@article{Folacci_2019b,
  title = {Regge pole description of scattering of gravitational waves by a {Schwarzschild} black hole},
  author = {Folacci, Antoine and Hadj, Mohamed Ould El},
  journal = {Phys. Rev. D},
  volume = {100},
  issue = {6},
  pages = {064009},
  numpages = {17},
  year = {2019},
  month = {Sep},
  publisher = {American Physical Society},
  doi = {10.1103/PhysRevD.100.064009},
  url = {https://link.aps.org/doi/10.1103/PhysRevD.100.064009}
}

@article{Andersson_1994,
doi = {10.1088/0264-9381/11/12/013},
url = {https://dx.doi.org/10.1088/0264-9381/11/12/013},
year = {1994},
month = {dec},
publisher = {},
volume = {11},
number = {12},
pages = {2991},
author = {Nils Andersson and Karl Erik Thylwe},
title = {Complex angular momentum approach to black-hole scattering},
journal = {Classical and Quantum Gravity},
}

@article{Bozza_2010,
  title={Gravitational lensing by black holes},
  author={Bozza, Valerio},
  journal={General Relativity and Gravitation},
  volume={42},
  number={9},
  pages={2269--2300},
  year={2010},
  publisher={Springer},
  url = {https://doi.org/10.1007/s10714-010-0988-2},
}

@article{Sanchez_1978b,
  title = {Elastic scattering of waves by a black hole},
  author = {S\'anchez, Norma},
  journal = {Phys. Rev. D},
  volume = {18},
  issue = {6},
  pages = {1798--1804},
  numpages = {0},
  year = {1978},
  month = {Sep},
  publisher = {American Physical Society},
  doi = {10.1103/PhysRevD.18.1798},
  url = {https://link.aps.org/doi/10.1103/PhysRevD.18.1798}
}

@article{Sanchez_1978a,
  title = {Absorption and emission spectra of a {Schwarzschild} black hole},
  author = {Sanchez, Norma},
  journal = {Phys. Rev. D},
  volume = {18},
  issue = {4},
  pages = {1030--1036},
  numpages = {0},
  year = {1978},
  month = {Aug},
  publisher = {American Physical Society},
  doi = {10.1103/PhysRevD.18.1030},
  url = {https://link.aps.org/doi/10.1103/PhysRevD.18.1030}
}

@article{Sanchez_1976,
    author = {Sanchez, Norma G.},
    title = {Scattering of scalar waves from a {Schwarzschild} black hole},
    journal = {Journal of Mathematical Physics},
    volume = {17},
    number = {5},
    pages = {688-692},
    year = {1976},
    month = {05},
    issn = {0022-2488},
    doi = {10.1063/1.522949},
}

@article{Handler_Matzner_1980,
  title = {Gravitational wave scattering},
  author = {Handler, F. A. and Matzner, Richard A.},
  journal = {Phys. Rev. D},
  volume = {22},
  issue = {10},
  pages = {2331--2348},
  numpages = {0},
  year = {1980},
  month = {Nov},
  publisher = {American Physical Society},
  doi = {10.1103/PhysRevD.22.2331},
  url = {https://link.aps.org/doi/10.1103/PhysRevD.22.2331}
}

@article{Fabbri_1975,
  title = {Scattering and absorption of electromagnetic waves by a {Schwarzschild} black hole},
  author = {Fabbri, R.},
  journal = {Phys. Rev. D},
  volume = {12},
  issue = {4},
  pages = {933--942},
  numpages = {0},
  year = {1975},
  month = {Aug},
  publisher = {American Physical Society},
  doi = {10.1103/PhysRevD.12.933},
  url = {https://link.aps.org/doi/10.1103/PhysRevD.12.933}
}

@article{Bao_2022,
    author = "Bao, ShouShan and Hou, Shaoqi and Zhang, Hong",
    title = "{Searching for wormholes with gravitational wave scattering}",
    doi = "10.1140/epjc/s10052-023-11281-9",
    journal = "Eur. Phys. J. C",
    volume = "83",
    number = "2",
    pages = "127",
    year = "2023"
}

@article{Chan_2025,
  title = {Lensing and wave optics in the strong field of a black hole},
  author = {Chan, Juno C. L. and others},
  journal = {Phys. Rev. D},
  volume = {112},
  issue = {6},
  pages = {064009},
  numpages = {16},
  year = {2025},
  month = {Sep},
  publisher = {American Physical Society},
  doi = {10.1103/6h6r-46cd},
  url = {https://link.aps.org/doi/10.1103/6h6r-46cd}
}

@article{Saketh_2025,
  title={Strong-field Gravitational Wave Lensing in the {Kerr} Background},
  author={Saketh, MVS and Ghosh, Rajes and Mishra, Anuj},
  journal={arXiv preprint arXiv:2511.23110},
  year={2025}
}

@article{Zhang_Fan_2021,
author = {H. Zhang and X. Fan},
title = {{Poisson-Arago} spot for gravitational waves},
journal = {Sci. China Phys. Mech. Astron.},
  volume = {64},
  issue = {120462},
  year = {2021},
  url = {https://doi.org/10.1007/s11433-021-1764-y}
}

@article{Hou_2019_b,
  author        = {Hou, Shaoqi and others},
  title         = {{Gravitational Wave Interference via Gravitational Lensing: Measurements of Luminosity Distance, Lens Mass, and Cosmological Parameters}},
  journal       = {Phys. Rev. D},
  volume        = {101},
  year          = {2020},
  pages         = {064011},
  doi           = {10.1103/PhysRevD.101.064011},
}

@misc{NIST:DLMF,
key = "{\relax DLMF}",
title = "{\it NIST Digital Library of Mathematical Functions}",
url = "https://dlmf.nist.gov/",}

@article{Li_2025,
doi = {10.1088/1475-7516/2025/01/092},
url = {https://dx.doi.org/10.1088/1475-7516/2025/01/092},
year = {2025},
month = {jan},
publisher = {IOP Publishing},
volume = {2025},
number = {01},
pages = {092},
author = {Li, Zhao and others},
title = {{Schwarzschild} lensing from geodesic deviation},
journal = {Journal of Cosmology and Astroparticle Physics},
}

@article{Isaacson_1968,
  title = {{Gravitational Radiation in the Limit of High Frequency. I. The Linear Approximation and Geometrical Optics}},
  author = {Isaacson, Richard A.},
  journal = {Phys. Rev.},
  volume = {166},
  issue = {5},
  pages = {1263--1271},
  numpages = {0},
  year = {1968},
  month = {Feb},
  publisher = {American Physical Society},
  doi = {10.1103/PhysRev.166.1263},
  url = {https://link.aps.org/doi/10.1103/PhysRev.166.1263}
}

@article{Hou_2019,
  title = {Gravitational lensing of gravitational waves: Rotation of polarization plane},
  author = {Hou, Shaoqi and Fan, XiLong and Zhu, ZongHong},
  journal = {Phys. Rev. D},
  volume = {100},
  issue = {6},
  pages = {064028},
  numpages = {8},
  year = {2019},
  month = {Sep},
  publisher = {American Physical Society},
  doi = {10.1103/PhysRevD.100.064028},
  url = {https://link.aps.org/doi/10.1103/PhysRevD.100.064028}
}

@article{Yuan_2025,
  title={Bayesian Analysis of Wave-Optics Gravitationally Lensed Massive Black Hole Binaries with Space-Based Gravitational Wave Detector},
  author={Yuan, Yong and others},
  journal={arXiv preprint arXiv:2509.01888},
  year={2025}
}

@article{Li_2025_b,
  title = {Rigorous calculation of scalar scattering in the {Schwarzschild} background: The convergence of the partial-wave series and the {Poisson} spot},
  author = {Li, Zhao and Zhao, Wen},
  journal = {Phys. Rev. D},
  volume = {112},
  issue = {8},
  pages = {083030},
  numpages = {17},
  year = {2025},
  month = {Oct},
  publisher = {American Physical Society},
  doi = {10.1103/xsxj-9vdw},
  url = {https://link.aps.org/doi/10.1103/xsxj-9vdw}
}

@article{Kinnersley_1969,
    author = {Kinnersley, William},
    title = {{Type D Vacuum Metrics}},
    journal = {Journal of Mathematical Physics},
    volume = {10},
    number = {7},
    pages = {1195-1203},
    year = {1969},
    month = {07},
    issn = {0022-2488},
    doi = {10.1063/1.1664958},
    url = {https://doi.org/10.1063/1.1664958}
}

@article{Jhingan_2003,
  title = {Improvement on the metric reconstruction scheme in the Regge-Wheeler-Zerilli formalism},
  author = {Jhingan, Sanjay and Tanaka, Takahiro},
  journal = {Phys. Rev. D},
  volume = {67},
  issue = {10},
  pages = {104018},
  numpages = {9},
  year = {2003},
  month = {May},
  publisher = {American Physical Society},
  doi = {10.1103/PhysRevD.67.104018},
  url = {https://link.aps.org/doi/10.1103/PhysRevD.67.104018}
}

@article{Hyun_2019,
  title = {Exact amplitudes of six polarization modes for gravitational waves},
  author = {Hyun, Younghwan and Kim, Yoonbai and Lee, Seokcheon},
  journal = {Phys. Rev. D},
  volume = {99},
  issue = {12},
  pages = {124002},
  numpages = {15},
  year = {2019},
  month = {Jun},
  publisher = {American Physical Society},
  doi = {10.1103/PhysRevD.99.124002},
  url = {https://link.aps.org/doi/10.1103/PhysRevD.99.124002}
}

@article{Shan_2025,
  title={An interference-based method for the detection of strongly lensed gravitational waves},
  author={Shan, Xikai and others},
  journal={Nature Astronomy},
  volume={9},
  pages={916–924},
  year={2025},
  publisher={Nature Publishing Group UK London}
}

@article{Martel_2005,
  title = {Gravitational perturbations of the {Schwarzschild} spacetime: A practical covariant and gauge-invariant formalism},
  author = {Martel, Karl and Poisson, Eric},
  journal = {Phys. Rev. D},
  volume = {71},
  issue = {10},
  pages = {104003},
  numpages = {13},
  year = {2005},
  month = {May},
  publisher = {American Physical Society},
  doi = {10.1103/PhysRevD.71.104003},
  url = {https://link.aps.org/doi/10.1103/PhysRevD.71.104003}
}

@misc{LVK_lensing,
title={GWTC-4.0: Searches for Gravitational-Wave Lensing Signatures}, 
author={The LIGO Scientific Collaboration and The Virgo Collaboration and The KAGRA Collaboration},
year={2025},
eprint={2512.16347},
archivePrefix={arXiv},
primaryClass={gr-qc},
url={https://arxiv.org/abs/2512.16347}, 
}

@misc{Chan_2025_b,
title={Discovering gravitational waveform distortions from lensing: a deep dive into GW231123}, 
author={Juno C. L. Chan and others},
year={2025},
eprint={2512.16916},
archivePrefix={arXiv},
primaryClass={astro-ph.CO},
url={https://arxiv.org/abs/2512.16916}, 
}

@misc{Goyal_2025,
title={Across the Universe: GW231123 as a magnified and diffracted black hole merger}, 
author={Srashti Goyal and Hector Villarrubia-Rojo and Miguel Zumalacarregui},
year={2025},
eprint={2512.17631},
archivePrefix={arXiv},
primaryClass={astro-ph.GA},
url={https://arxiv.org/abs/2512.17631}, 
}

@misc{Chakraborty_2025,
title={{The First Model-Independent Upper Bound on Micro-lensing Signature of the Highest Mass Binary Black Hole Event GW231123}}, 
author={Aniruddha Chakraborty and Suvodip Mukherjee},
year={2025},
eprint={2512.19077},
archivePrefix={arXiv},
primaryClass={gr-qc},
url={https://arxiv.org/abs/2512.19077}, 
}

@misc{Shan_2025_b,
title={{GW231123: A Case for Binary Microlensing in a Strong Lensing Field}}, 
author={Xikai Shan and Huan Yang and Shude Mao},
year={2025},
eprint={2512.19118},
archivePrefix={arXiv},
primaryClass={astro-ph.GA},
url={https://arxiv.org/abs/2512.19118}, 
}

\end{document}